\documentclass{IEEEtran}

\usepackage{algpseudocode}
\usepackage[dvips]{graphicx}
\usepackage{latexsym}
\usepackage{amssymb}
\usepackage{amsmath}
\usepackage{multirow}
\usepackage{xcolor}
\usepackage{lipsum}
\usepackage{multicol}
\usepackage{enumerate}
\usepackage{algorithm}
\usepackage{algorithmicx}
\usepackage{algpseudocode}
\usepackage{amsmath}
\usepackage{graphicx}
\usepackage{subfig}
\usepackage[utf8]{inputenc}
\usepackage{float}
\usepackage{bm}
\usepackage{cite}
\usepackage{caption}
\usepackage{booktabs}
\definecolor{mycolor}{rgb}{0,0,1}
\definecolor{my}{rgb}{0,0,0}
\definecolor{my1}{rgb}{0,0,0}

\begin{document}

\title{\textcolor{my}{A New Solution} for MU-MISO Symbol-Level Precoding: Extrapolation and Deep Unfolding}

\author{\textcolor{my}{Mu Liang, Ang Li,~\IEEEmembership{Senior Member,~IEEE},\\Xiaoyan Hu,~\IEEEmembership{Member,~IEEE}, and Christos Masouros,~\IEEEmembership{Fellow,~IEEE}}

	}

\maketitle

\begin{abstract}
Constructive interference (CI) precoding, which converts the harmful multi-user interference into beneficial signals, is a promising and efficient interference management scheme in multi-antenna communication systems. However, CI-based symbol-level precoding (SLP) experiences high computational complexity as the number of symbol slots increases within a transmission block, rendering it unaffordable in practical communication systems. In this paper, we propose a symbol-level extrapolation (SLE) strategy \textcolor{my}{to extrapolate the precoding matrix} by leveraging the relationship between different symbol slots within in a transmission block, during which the channel state information (CSI) remains constant, where we design a closed-form iterative algorithm based on SLE for both PSK and
QAM modulation. In order to further reduce the computational complexity, a sub-optimal closed-form solution based on SLE is further developed for PSK and QAM, respectively. 
Moreover, we design an unsupervised SLE-based neural network (SLE-Net) to unfold the proposed iterative algorithm, which helps enhance the interpretability of the neural network. By carefully designing the loss function of the SLE-Net, the time-complexity \textcolor{my}{of the network} can be reduced effectively.
Extensive simulation results illustrate that the proposed algorithms can dramatically reduce the computational complexity and time complexity with only marginal performance loss, compared with the conventional SLP design methods.

\end{abstract}

\section{Introduction}
\IEEEPARstart{M}{ulti}-antenna communication has the potential to offer significant multiplexing and array gains, which can meet the spectral efficiency and energy efficiency requirements for current and future wireless communication systems\cite{rusek2012scaling}. Due to the existence of inter-stream and inter-user interference, careful management of interference is crucial for realizing the benefits of multi-antenna communication systems\cite{marzetta2010noncooperative}. In the downlink transmission, precoding, which is an effective technique for managing interference, has attracted extensive attention\cite{zheng2003diversity}.\par

%开始介绍非线性预编码
Among various precoding approaches, the \textcolor{my}{capacity-achieving} dirty-paper coding (DPC) scheme is proposed in\cite{costa1983writing} by pre-subtracting the interference prior to transmission. However, DPC is difficult to implement in practical systems due to the assumption of infinite codebook length. Tomlinson-Harashima precoding (THP) \cite{sun2013quantized} imposes an integer offset at the transmitter, while a modulo operation is required for the received signal. In \cite{hochwald2005vector}, vector perturbation (VP) precoding is propopsed, which selects an appropriate perturbation vector to perturb the data symbols via the sphere encoding algorithm and \textcolor{my}{a} modulo operation is also required at the receiver side. However, all of the above \textcolor{my}{precoding} methods demand significant computational resources due to their nonlinear design nature, and are too complicated to implement.
%开始介绍线性预编码
In order to find a balance between performance and complexity, linear precoding methods are proposed and have received more attention\cite{lo1999maximum,haustein2002performance,peel2005vector}. \textcolor{my}{Maximum-ratio} transmission (MRT) precoding is the simplest strategy that maximizes the received signal-to-noise ratio (SNR) \cite{lo1999maximum}, while the capability of MRT to handle \textcolor{my}{inter-user} interference is limited. In \cite{haustein2002performance}, zero-forcing (ZF) precoding that employs the channel inversion to fully eliminate \textcolor{my}{mutual} interference is proposed. Additionally, a regularized ZF (RZF) scheme is proposed in \cite{peel2005vector} to alleviate the noise amplification effect of ZF precoding to further improve the communication performance.\par

On the other hand, precoding algorithms assisted by optimization theory have received increasing attention.
%功率最小化的描述 
The well-known power minimization problem (PM) aims to minimize the total transmit power subject to the \textcolor{my}{signal-to-interference-plus-noise} ratio (SINR) target for each user. This problem can be solved using various approaches, including the uplink-downlink duality\cite{visotsky1999optimum}, conic programming\cite{wiesel2005linear}, semidefinite relaxation method (SDR)\cite{bengtsson1999optimal}, iteartive algorithm, etc.
%SINR balance描述
Another popular form of optimization-based schemes is known as SINR balancing (SB) problem, which aims to maximize the minimum SINR of the users subject to a total power constraint. The SINR balancing problem (SB) is proved to be the inverse problem of the power minimization problem. Several schemes have been proposed to address the SB precoding problem, including bisection search \cite{wiesel2005linear} and the iterative algorithm \cite{schubert2004solution}. \textcolor{my}{In addition to the sum-power constraint, the consideration of a per-antenna power constraint may be more practical, because} antenna has its dedicated power amplifier. In \cite{wen2023interference}, the authors employ the primal-dual \textcolor{my}{interior-point} method (IPM) to solve the SB problem under the \textcolor{my}{per-antenna power constraint} (PAPC). In \cite{7422140}, the authors consider signal-to-leakage-plus-noise ratio (SLNR) \textcolor{my}{maximization} for multi-antenna \textcolor{my}{downlink} under the PAPC based on \textcolor{my}{the} gradient projection \textcolor{my}{method}. \par

%开始介绍slp
It is important to note that, all of the above precoding design methods view interference as harmful and attempt to eliminate the interference, while the interference can in fact be beneficial and further exploited on a symbol level via the concept of symbol-level precoding \cite{masouros2013known}. The idea of exploiting interference was first introduced in \cite{4289141}, where instantaneous interference was classified into constructive and destructive. Some early approaches to exploit constructive interference (CI) were first proposed in \cite{4801492}. In\cite{masouros2010correlation}, the authors proposed a correlation rotation linear precoding scheme to translate destructive interference into constructive. Subsequently, optimization-based interference exploitation methods have been proposed in \cite{masouros2015exploiting} and \cite{alodeh2015constructive} to expand the region of CI by relaxing the strict phase-rotation requirement to achieve a better performance.\par

%挑战，调研一些降低复杂度的方法以及deepunfolding模型驱动的方法
For SLP, the precoding matrix is dependent on both the channel state information (CSI) and data symbols, which means that the SLP methods must solve a constrained optimization problem at each symbol slot to fully exploit the benefits of CI. Within a transmission block, when the CSI remains constant while the data symbols change \textcolor{my}{per slot,} recalculating the precoding matrix becomes inevitable, which results in huge computational burden for SLP compared to traditional \textcolor{my}{precoding approaches}. Therefore, the primary challenge to implement SLP in practical communication systems is to reduce the substantial computational complexity. In recent years, low-complexity SLP methods have attracted increasing attention. 
In \cite{li2018interference}, the authors proposed an iterative closed-form scheme to calculate the optimal precoding for SLP. 
Compared to directly solving optimization problems, the proposed iterative algorithm offers substantially reduced computational complexity. In \cite{yang2022low}, the authors reformulated the \textcolor{my1}{CI-based} PM problem and showed that the optimization problem was able to be decomposed into several parallel sub-problems, leading to a substantial reduction in computational complexity. 
In \cite{xiao2022low}, the authors divided the users into groups according to their channel correlation, and then designed the precoding matrix where the interference is exploited within each user group while eliminate between different groups to reduce the \textcolor{my}{complexity}, which however results in a sub-optimal performance. In \cite{li2022practical}, a CI-based block-level precoding (BLP) was proposed, in which a same precoding matrix is applied to a block of symbol slots within one transmission block to reduce the calculation complexity. BLP offers some additional performance benefits when the length of the considered block is short, \textcolor{my}{while} its performance will deteriorate as the block length increases. Therefore, it requires further explorations of low-complexity SLP designs.

\par

%hu2020iterative 需要修改，照抄过来的 Iterative algorithm induced deep-unfolding neural networks: Precoding design for multiuser MIMO systems
Recently, many studies have developed learning-based algorithms to solve the computationally intensive and time sensitive signal processing tasks in communication systems. In \cite{lei2021ci}, the authors introduced a model-driven neural network for PSK-based SLP design, reducing the complexity of obtaining the precoding matrix. Nevertheless, it didn't adequately consider the network's generalizability, thus constraining its practical applicability in practical environments. 
In \cite{zhang2022deep}, the authors proposed a learning-based framework for low-complexity multi-user MIMO precoding design, employing a neural network to fit the mapping function between channel coefficients and the precoding matrix. However, it remains fundamentally a data-driven neural network, and the interpretability of the neural network requires further enhancement.
Deep unfolding has become \textcolor{my}{an} increasingly popular method that transforms iterative algorithms into a layer-wise structure, resembling a neural network \cite{balatsoukas2019deep}. The deep unfolding enhances the network's interpretability by transforming the network from data-driven to model-driven. This method has a wide range of applications in communications, such as detection and coding \cite{cammerer2017scaling,samuel2019learning,he2018model}. 
For example, in \cite{mohammad2023unsupervised}, the authors unfolded a power minimization SLP formulation based on the \textcolor{my}{IPM} proximal barrier function and proposed a robust precoding matrix design method. In \cite{hu2020iterative}, the author proposed a deep-unfolding framework to solve the sum-rate maximization problem for precoding design in MU-MIMO system and reduced the complexity compared to the conventional method.

\par
On the other hand, extrapolation is a method to estimate or predict unknown data by utilizing information from known data \textcolor{my1}{and their correlation}, which finds extensive applications in channel estimation and prediction \textcolor{my}{in recent years} \cite{deepe,alrabeiah2019deep}. In \cite{deepe}, \textcolor{my}{channel extrapolation} is proposed for obtaining the CSI of different antenna patterns in the pattern reconfigurable communication system, where reducing the pilot overheads \textcolor{my}{is the main challenge}. 
In \cite{alrabeiah2019deep}, \textcolor{my}{channel extrapolation} is proposed for obtaining the downlink channel coefficients from the uplink channel coefficients \textcolor{my}{in a FDD MIMO system}, where \textcolor{my}{the pilot overheads are greatly reduced}.
\par

Essentially, the reason why SLP is computationally expensive is that SLP methods must solve a constrained optimization problem for each data symbol slot. Motivated by the extrapolation method, in this paper, we propose an iterative symbol-level extrapolation (SLE) method to extrapolate the SLP precoding matrix for different data symbol slots to reduce the complexity of CI-based SLP design. Furthermore, based on the SLE, we unfold the iteration algorithm into a unique layer-wise structure and derive a SLE-Net. For clarity, we summarize the contributions of this paper as:
\begin{enumerate}
	\item We propose a novel low-complexity SLE-based SLP solution. Within a transmission block, we exploit the relationships between different symbol slots to extrapolate the symbol-scaling factors of data symbols used in SLP to reduce the complexity.
	\item By describing the CI-SLP problem using symbol-scaling metric, we simplify optimization problem into a \textcolor{my1}{convex} problem. Based on the simplified problem, the efficient SLE-based iterative algorithm is developed to solve the SLP problem for PSK modulation and QAM modulation, respectively. Moreover, we also derive a sub-optimal closed-form algorithm, to further reduce the computational complexity.
	\item We utilize the unfolding approach and design an unsupervised SLE-Net by expanding the SLE-based iterative algorithm. Based on the simplified formulation of the optimization problem for SLP, we design a unique loss function for the SLE-Net.
	\item  We study the computational costs of the proposed schemes in terms of the required number of multiplications.
\end{enumerate}

Numerical results show that the proposed algorithm can reduce complexity. Compared to traditional SLP \textcolor{my1}{algorithms}, the proposed SLE-based method can reduce computational complexity without significant performance losses for \textcolor{my1}{both} PSK modulation and QAM modulation. At the same time, the proposed SLP-Net, with comparable time complexity to the proposed SLE-based sub-optimal closed-form algorithm, exhibits marginal performance loss \textcolor{my}{to the optimal solution}\textcolor{my1}{, exhibiting the superiority of the SLP-Net.}.

\emph{Notations:}
$\alpha$, $\bm{\alpha}$, $\bf{A}$ denote scalar, vector and matrix, respectively. $vec\left ( \cdot  \right )$ denotes the vector form of a matrix, $\Re(\cdot)$ and $\Im(\cdot)$ extract the real and imaginary parts of the argument, respectively.
$\left \| \bf{\cdot} \right \| _{2} $ denotes the Euclidean norm. $(\cdot)^H$ and $(\cdot)^{T}$ denotes the conjugate transposition and transposition. $diag(\cdot)$ is the transformation of a column vector into a diagonal matrix. $card\left|\cdot\right|$ denotes the cardinality of a set, $\otimes$ represents the Kronecker product, and $\odot$ represents the Hadamard product operation. $\mathbb{R}$ and $\mathbb{C}$ denote the set of the real numbers and complex numbers, respectively. \textcolor{my}{${\bf I}_{K}$ denotes the $K\times K$ identity matrix.}\par

\section{System Model And Constructive Inference} 
\subsection{System Model}

We consider a multi-user MISO system in the downlink equipped with $M_{\mathrm{t}}$ transmit antennas \textcolor{my}{serving} $K$ single-antenna users. We consider \textcolor{my}{the scenario of} a limited number of scattering clusters between the transmitter and each receiver \cite{zhang2021training}, and therefore channel coefficients for the $k$-th user, denoted as ${\bf h}_{k}\in \mathbb{C}^{M_{\mathrm{t}}\times 1} $, can be modeled as
\footnote{\textcolor{my1}{The proposed extrapolation method is channel independent, and applies to other channel models as well.}}

\begin{equation}
	{\bf h}_{k}=\sum_{l=1}^{L_{k}} \alpha _{k,l} {\mathbf{a}}_{\mathrm{t}}(\theta _{k,l}),
	\label{1}
\end{equation}
where $L_{k}$ is the total number of the propagation paths between the transmitter and the $k$-th receiver, $\alpha _{k,l}$ is the complex gain of the $k$-th user’s $l$-th path following the complex Gaussian distribution. ${\mathbf{a}}_{\mathrm{t}}(\theta _{k,l})\in \mathbb{C}^{M_{\mathrm{t}}\times1}$ is \textcolor{my1}{referred to} the antenna array steering vector. When uniform linear array (ULA) is used, ${\mathbf{a}}_{\mathrm{t}}(\theta _{k,l})$ can be expressed as

\begin{equation}
	{\mathbf{a}}_{\mathrm{t}}(\theta _{k,l}) = \frac{1}{\sqrt{M_{\mathrm{t}}}}\left[1,e^{j\theta_{k,l}},\cdots,e^{j(M_{\mathrm{t}}-1)\theta_{k,l}}\right]^{T}.
	\label{2}
\end{equation}
In (\ref{2}), $\theta _{k,l}$ denotes the normalized angle of departure (AoD) of the $k$-th user's $l$-th path, and the relationship with the physical AoD $\phi_{k,l}$ can be expressed as
\begin{equation}
	\theta _{k,l} = \frac{2 \pi d \sin(\phi_{k,l}) }{\lambda},
\end{equation}
where $d$ is the spacing between two adjacent antenna elements at the transmitter and $\lambda$ is the wavelength. Moreover, by assuming that $\phi _{k,l}\in [0,2\pi )$ and $d =\frac{\lambda}{2} $, we have $\theta_{k,l}\in [-\pi,\pi)$. To better describe the channel in a time-varying environment, we divide the normalized AoD into $N$ grids, as shown in the following \textcolor{my}{formula:}

\begin{equation}
	\bar{\theta}_{n} = \frac{2\pi (n-1)}{N}+\frac{\pi }{N},\forall n=1,\cdots ,N. 
\end{equation}

\textcolor{my}{In this work, we consider} the dynamic channel model of the time-varying AoDs and the complex path gains\textcolor{my}{, where} normalized AoD follows the discrete state Markov processing. \textcolor{my}{Thus, the} angle from the $(c-1)$-th transmission block to the $c$-th transmission block is determined by the transition probability function in \cite{zhang2021training}.
The complex path gain $\alpha_{k,l}$ is assumed to follow an auto-regressive process, given by
\begin{equation}
	\alpha_{k,l}^{c} =  p_{k}\alpha_{k,l}^{c-1} + \eta_{k,l}^{c},
\end{equation}
where $p_{k}\in [0,1]$ is the correlation coefficient, \textcolor{my1}{$\eta_{k,l}^{c}$ is a random variable used to simulate an auto-regressive process}, $\eta_{k,l}^{c}$ and $\alpha_{k,l}$ are assumed to follow the distribution of $\mathcal{CN}(0,(1-p_{k}^2)\sigma ^{2})$ and $\mathcal{CN}(0,\sigma ^{2})$, \textcolor{my1}{$\sigma ^{2}$ denotes the variance of the distribution}.\par

Within a transmission block, the \textcolor{my}{data symbol vector in} the $n$-th \textcolor{my}{slot ${\bf s}^{(n)} = \left[s_{1}^{(n)},s_{2}^{(n)},\cdots,s_{K}^{(n)}\right]^{T} \in \mathbb{C}^{K \times 1}$} \textcolor{my1}{is} \textcolor{my}{drawn from a normalized constellation}, and the received signal at the $k$-th user in a downlink MU-MISO system can be expressed as
\textcolor{my}{
\begin{equation}
	y_{k}^{(n)} =  {\bf h}_{k}^{T}{\bf W}^{(n)}{\bf s}^{(n)} + z_{k}^{(n)},
	\label{6}
\end{equation}
}where \textcolor{my}{${\bf W}^{(n)} \in \mathbb{C}^{M_{\mathrm{t}}\times K}$} is the precoding matrix for the $n$-th symbol slot, \textcolor{my}{and} $z_{k}^{(n)}$ is the additive Gaussian noise at the receiver following the standard complex Gaussian distribution. 

\subsection{Constructive Interference for PSK Modulation}

CI refers to the interference that pushes the received signals away from all of their detection thresholds, \textcolor{my}{leading to higher probability for correct signal detection} \cite{li2020tutorial}. Therefore, based on this definition, the constructive region is the complex plane within which the interference is able to push the received signal, making it farther away from the decision boundaries. In this paper, we choose the symbol-scaling metric in \cite{li2020interference} to describe the CI \textcolor{my1}{constrains}, which is applicable to both PSK and QAM \textcolor{my1}{modulations}. The symbol-scaling CI metric decomposes the given constellation point along the directions of the vectors that define the detection thresholds.\par 

\begin{figure}
	\centering
	\includegraphics[width=0.4\textwidth]{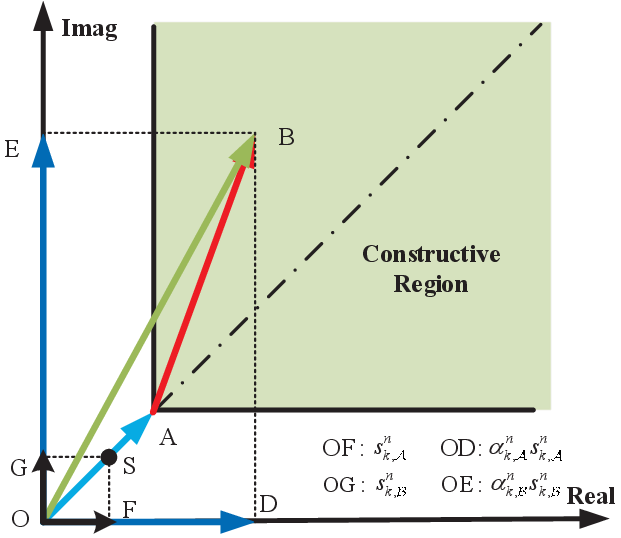}
	\captionsetup{labelformat=default,labelsep=space}
	\caption{CI regions for generic $\mathcal{M}$-PSK}
	\label{psk_ci}
\end{figure}

\textcolor{my}{As an example}, we discuss the \textcolor{my1}{CI region} of QPSK modulation, and the CI constraints are illustrated in Fig. \ref{psk_ci}. \textcolor{my}{Without loss of generality, we denote $\vec{OS} =s_{k}^{(n)}$ as a nominal constellation point that is the intended data symbol for user $k$ in the $n$-th slot. We further denote $\vec{OB} ={\bf h}_{k}^{T}{\bf W}^{(n)}{\bf s}^{(n)}$ as the received signal for user $k$ in the $n$-th slot excluding noise.} \textcolor{my1}{Based on the geometry in Fig. 1, we are able to} decompose the constellation symbol along the detection thresholds \textcolor{my}{as:}
\begin{equation}
	\vec{OS} = \vec{OF} + \vec{OG}  \Rightarrow s_{k}^{(n)}=s_{k,\mathcal{A}}^{(n)}+s_{k,\mathcal{B}}^{(n)}.
	\label{7}
\end{equation}

For generic $\mathcal{M}$-PSK constellations, each data symbol can be expressed as
\begin{equation}
	s_{\mathrm{c}} = e^{j\left[\frac{2\pi}{\mathcal{M}}(c-1)+\frac{\pi}{4}\right]}, c\in \{1,2,\cdots,\mathcal{M}\},
\end{equation}
where $s_{{c}}$ denotes the $c$-th constellation point. For the $k$-th data symbol \textcolor{my}{in} the $n$-th symbol slot in one transmission block, the detection thresholds $s_{k,\mathcal{A}}^{(n)}$ and $s_{k,\mathcal{B}}^{(n)}$ of constellation point $s_{{k}}^{(n)}$ can be expressed as
\begin{equation}
	s_{k,\mathcal{A}}^{(n)} = \frac{s_{{k}}^{(n)}e^{j\frac{\pi}{\mathcal{M}}}}{2\cos(\frac{\pi}{\mathcal{M}})},
	\label{9}
\end{equation}
\begin{equation}
	s_{k,\mathcal{B}}^{(n)} = \frac{s_{{k}}^{(n)}e^{-j\frac{\pi}{\mathcal{M}}}}{2\cos(\frac{\pi}{\mathcal{M}})}.
	\label{10}
\end{equation}

Following a similar approach to (\ref{7}), we decompose the noiseless received signal for user $k$ along the detection thresholds, which is shown as
\textcolor{my}{
\begin{equation}
	\vec{OB} = \vec{OD} + \vec{OE}  \Rightarrow {\bf h}_{k}^{T}{\bf W}^{(n)}{\bf s}^{(n)}=\alpha _{k,\mathcal{A}}^{(n)}s_{k,\mathcal{A}}^{(n)}+\alpha _{k,\mathcal{B}}^{(n)}s_{k,\mathcal{B}}^{(n)},
	\label{11}
\end{equation}
}where $\alpha _{k,\mathcal{A}}^{(n)}>0$ and $\alpha _{k,\mathcal{B}}^{(n)}>0$ represent the symbol-scaling factors along the detection thresholds for the $k$-th constellation of the $n$-th symbol slot. The \textcolor{my1}{set} $\mathcal{O}^{(n)}$ and $\mathcal{I}^{(n)}$ \textcolor{my1}{consists of the} symbol-scaling \textcolor{my1}{factors} for the $n$-th symbol slot that can be exploited interference and cannot be exploited interference, respectively. \textcolor{my1}{For PSK modulation, all constellation points of PSK modulation can fully leverage the advantages of CI. Therefore, $\mathcal{O}^{(n)}$ and $\mathcal{I}^{(n)}$ can be expressed as
\begin{equation}
	\mathcal{I}^{(n)}=\emptyset,
\end{equation}
\begin{equation}
	\mathcal{O}^{(n)}=  \left\{ {\alpha _{1,\cal A}^{(n)} ,{\alpha _{1,\cal B}^{(n)}},{\alpha _{2,\cal A}^{(n)}},{\alpha _{2,\cal B}^{(n)}}, \cdots ,{\alpha _{K,\cal A}^{(n)}},{\alpha _{K,\cal B}^{(n)}}} \right\}. 
\end{equation}}
\subsection{Constructive Interference for QAM Modulation}

\begin{figure}
	\centering
	\includegraphics[width=0.38\textwidth]{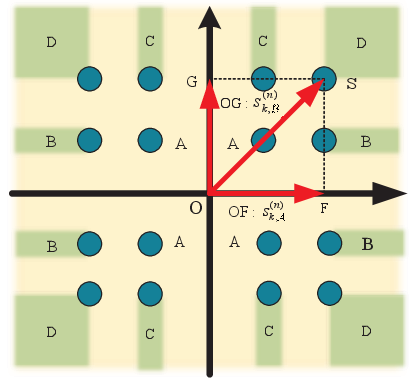}
	\captionsetup{labelformat=default,labelsep=space}
	\caption{CI regions for 16QAM}
	\label{qam_ci}
\end{figure}

For generic $\mathcal{M}$-QAM modulation, we use 16QAM as an example to explain the impact of CI, as illustrated in Fig. \ref{qam_ci}. There are four types of constellation points for QAM modulation, denoted as `A', `B', `C', and `D'. The interference on the inner constellation points of type \textcolor{my1}{`A'} is only destructive, and CI exists for the real part of the constellation points type \textcolor{my1}{`B'} and imaginary \textcolor{my1}{parts} of type \textcolor{my1}{`C'}, while both the real and imaginary part of the constellation points type \textcolor{my1}{`D'} can be exploited. To be more specific, we decompose the constellation symbol along the detection thresholds, as follows
\begin{equation}
	\vec{OS} = \vec{OF} + \vec{OG}  \Rightarrow s_{k}^{(n)}=s_{k,\mathcal{A}}^{(n)}+s_{k,\mathcal{B}}^{(n)},
	\label{12}
\end{equation}
For the $k$-th data symbol of the $n$-th symbol slot in one transmission block, $s_{k,\mathcal{A}}^{(n)}$ and $s_{k,\mathcal{B}}^{(n)}$ can be expressed as
\textcolor{my}{
\begin{equation}
	s_{k,\mathcal{A}}^{(n)} =\Re \left({s_{k}^{(n)}}\right), s_{k,\mathcal{B}}^{(n)} =j \cdot \Im \left(s_{k}^{(n)}\right).
	\label{13}
\end{equation}
}\textcolor{my1}{For QAM modulation, only the outer constellation points can exploit CI, thus their symbol-scaling factors belong to $\mathcal{O}^{(n)}$, while the inner constellation points cannot exploit CI, and their symbol-scaling factors belong to $\mathcal{I}^{(n)}$. Therefore, $\mathcal{O}^{(n)}$ and $\mathcal{I}^{(n)}$ for the $n$-th symbol slot can be expressed as
\begin{equation}
	{\cal O}^{(n)} \cup {\cal I}^{(n)} = \left\{ {\alpha _{1,\cal A}^{(n)} ,{\alpha _{1,\cal B}^{(n)}},{\alpha _{2,\cal A}^{(n)}},{\alpha _{2,\cal B}^{(n)}}, \cdots ,{\alpha _{K,\cal A}^{(n)}},{\alpha _{K,\cal B}^{(n)}}} \right\},
	\label{38}
\end{equation}
and 
\begin{equation}
	card\left| {\cal O}^{(n)} \right| + card\left| {\cal I}^{(n)} \right| = 2K.
	\label{39}
\end{equation}}
\section{\textcolor{my}{Proposed} Symbol-Level Extrapolation Precoding Design} 
For $\mathcal{M}$-PSK and $\mathcal{M}$-QAM modulation, by leveraging the relationships of data symbols in different symbol slots within one transmission block, we propose an iterative SLE algorithm and a sub-optimal algorithm to reduce the computational complexity \textcolor{my}{of conventional SLP}.
Within one transmission block, the CSI is constant, while the precoding matrix for SLP is different between symbol slots, resulting \textcolor{my1}{in} significant computational complexity in SLP design. To address this issue of SLP, we leverage the relationships between different symbol \textcolor{my}{slots}, and without loss of generality we describe the extrapolation of the precoding matrix from the $n$-th symbol slot to the $m$-th symbol slot.\par

The data symbols for the $m$-th symbol slot and the $n$-th symbol slot are denoted as ${\bf s}^{(m)}\in \mathbb{C}^{K\times1}$ and ${\bf s}^{(n)}\in \mathbb{C}^{K\times1}$, respectively. The relationship between ${\bf s}^{(m)}$ and ${\bf s}^{(n)}$ can be defined as:
\begin{equation}
	{\bf s}^{(m)} = {\bf A}{\bf s}^{(n)},
	\label{14}
\end{equation}
where ${\bf A} = diag(\{\beta_{1}e^{j\cdot \varphi_{1}},\beta_{2}e^{j\cdot \varphi_{2}},\cdots ,\beta_{K}e^{j\cdot \varphi_{K}}\})\in \mathbb{C}^{K\times K} $ represents the diagonal transform matrix, $\varphi_{k}$ and $\beta_{k}$ denote the phase and amplitude difference between the $k$-th symbol in the $n$-th symbol slot and in the $m$-th symbol slot, respectively.

\subsection{The case of $\mathcal{M}$-PSK}

Based on the symbol-scaling metric in \cite{li2020interference}, the original optimization problem that maximizes the CI effect for the $n$-th symbol slot is give by
\textcolor{my1}{
\begin{equation}
	\begin{aligned}
		&\mathcal{P}_1: {\kern 3pt} \mathop {\max }\limits_{{{\bf{W}}^{(n)}}, {\kern 1pt} t} {\kern 3pt} t \\
		&{\kern 0pt} s. t. {\kern 10pt} {{\bf{h}}_k^T}{{\bf W}^{(n)}{\bf s}^{(n)}} = \alpha _{k,\mathcal{A}}^{(n)}s_{k,\mathcal{A}}^{(n)}+\alpha _{k,\mathcal{B}}^{(n)}s_{k,\mathcal{B}}^{(n)}, {\kern 3pt} \forall k \in {\cal K}, \\
		&{\kern 24pt} t \le \alpha_i^{\cal O}, {\kern 3pt} \forall \alpha_i^{\cal O} \in {\cal O}^{(n)},\\
		&{\kern 22pt} \left\| {{\bf{W}}^{(n)}{\bf s}^{(n)}} \right\|_2^2 \le {p_0},
		\label{p1}
	\end{aligned}
\end{equation}
}where $\mathcal{K}=\left\{ {1,2,\cdots,K}\right\}$, $\alpha_i^{\cal O}$ represents the $i$-th element in set ${\cal O}^{(n)}$, $p_{0}$ is the total power at the transmitter.

\par
$\mathcal{P}_1$ \textcolor{my}{is convex and} can be directly solved via convex optimization tools. In \cite{li2020interference}, the authors derive the optimal solution structure of the precoding matrix. For the $n$-th symbol slot, the precoding matrix can be expressed as
\textcolor{my}{
\begin{equation}
	\begin{aligned}
	{\bf W}^{(n)} &=\\&
	\frac{1}{K} \cdot {{\bf{H}}^H}{\left( {{\bf{H}}{{\bf{H}}^H}} \right)^{ - 1}}{\bf{U}}diag\left( {\bf \Omega}^{(n)}\right){\bf s}_{\bf E}^{(n)} \left({{{\bf{\hat s}}}^{(n)}}\right)^{T},
	\end{aligned}
	\label{18}
\end{equation}
}where \textcolor{my}{${\bf \Omega}^{(n)} = \left[\alpha _{1,\mathcal{A}}^{(n)},\alpha _{1,\mathcal{B}}^{(n)},\cdots,\alpha _{K,\mathcal{A}}^{(n)},\alpha _{K,\mathcal{B}}^{(n)}\right]^{T} \in \mathbb{R}^{2K\times1}$} is the symbol-scaling factors for the $n$-th symbol slot, ${\bf H}\in \mathbb{C}^{K\times M_{\mathrm{t}}}$ is the channel matrix, ${\bf{\hat s}}^{(n)} \in \mathbb{C}^{K\times1}$, ${\bf U} \in \mathbb{R}^{K\times2K}$ and ${\bf s}_{\bf E}^{(n)} \in \mathbb{C}^{2K\times1}$ are given by 
\textcolor{my}{
\begin{equation}
{\bf{\hat s}}^{(n)} = \left[\frac{1}{{\bf s}_{1}^{(n)}},\frac{1}{{\bf s}_{2}^{(n)}},\cdots,\frac{1}{{\bf s}_{K}^{(n)}}\right]^{T}\in \mathbb{C}^{K\times1},
\label{19}
\end{equation}
}
\begin{equation}
	{\bf U} = {\bf I}_{K} \otimes [1,1],
	\label{20}
\end{equation}
\begin{equation}
	{\bf s}_{\bf E}^{(n)} = {\left[ s_{1,\cal A}^{(n)},s_{1,\cal B}^{(n)} ,s_{2,\cal A}^{(n)},s_{2,\cal B}^{(n)} , \cdots ,s_{K,\cal A}^{(n)},s_{K,\cal B}^{(n)} \right]^T}.
	\label{21}
\end{equation}
When we obtain ${\bf \Omega}^{(n)}$, the optimal precoding matrix \textcolor{my}{can be directly obtained} by using the (\ref{18}). Therefore, the key issue is how to efficiently compute the symbol-scaling factors with low complexity. In \cite{li2020interference}, the authors utilize the Lagrangian dual and Karush-Kuhn-Tucker (KKT) conditions to transform the original problem $\mathcal{P}_1$ into \textcolor{my}{simpler} forms. \textcolor{my}{More specifically, the original} optimization problem can be simplified \textcolor{my}{into}
\textcolor{my1}{
\begin{equation}
	\begin{aligned}
		&\mathcal{P}_2: {\kern 3pt} \mathop {\min }\limits_{{{\bf \Omega}^{(n)}}, {\kern 1pt} t} {\kern 3pt} -t \\
		&{\kern 0pt} s. t. {\kern 10pt} {({\bf \Omega}^{(n)})}^T {\bf V}_{n} {\bf \Omega}^{(n)} - p_0 = 0,\\
		&{\kern 24pt} t - \alpha_i^{\cal O} \le 0, {\kern 3pt} \forall \alpha_i^{\cal O} \in {\cal O}^{(n)},\\		
	\end{aligned}
	\label{p2}
\end{equation}}where ${\bf V}_{n}\in \mathbb{R}^{2K\times2K}$ is \textcolor{my}{dependent on} both CSI and data symbols and can be expressed as
\begin{equation}
	{\bf V}_{n} = \Re\left({\bf T}_{n}\right),
	\label{25}
\end{equation}
\textcolor{my1}{\begin{equation}
	{\bf T}_{n} = {diag\left( ({\bf s}_{\bf E}^{(n)} )^{H} \right) {\bf Q}_{n} diag\left( {{\bf s}_{\bf E}^{(n)}}\right)},
	\label{24}
\end{equation}}
\begin{equation}
	{\bf Q}_n={{\bf{U}}^H}{{\left( {{\bf{H}}{{\bf{H}}^H}} \right)}^{ - 1}}{\bf{U}},
	\label{23}
\end{equation}
where the details can be found in \cite{li2020interference}.
\textcolor{my}{The optimization $\mathcal{P}_3$ on ${\bf \Omega}^{(m)}$} for the $m$-th symbol slot \textcolor{my}{is in the same form as $\mathcal{P}_2$ above.}
\par
Within one transmission block, we exploit the relationships between \textcolor{my}{the} $n$-th symbol slot and \textcolor{my}{the} $m$-th symbol slot to connect $\mathcal{P}_2$ and $\mathcal{P}_3$. \textcolor{my}{Based on} the diagonal transformation matrix ${\bf A}$ between the $n$-th symbol slot and $m$-th symbol slot \textcolor{my}{in} (\ref{14}),  \textcolor{my}{we} employ ${\bf A}$ to extrapolate ${\bf V}_{m}$ from the matrix ${\bf V}_{n}$, thereby establishing \textcolor{my}{the} connection \textcolor{my}{of the precoding matrix} between the two \textcolor{my}{symbol slots} to reduce the complexity. The relationship between ${\bf V}_{n}$ and ${\bf V}_{m}$ can be expressed as
\begin{equation}
	{\bf s}_{\bf E}^{(m)} = \left({{\bf A}\otimes {\bf I}_2 }\right){\bf s}_{\bf E}^{(n)},
	\label{27}
\end{equation}

\begin{equation}
	{\bf T}_{m} = \left({{\bf A}\otimes {\bf I}_2 }\right)^{H}{\bf T}_{n}\left({{\bf A}\otimes {\bf I}_2 }\right),
	\label{28}
\end{equation}

\begin{equation}
	{\bf V}_{m} = \Re\left({\bf T}_{m}\right).
	\label{29}
\end{equation}

Based on the above derivations and the specific solution structure of \textcolor{my1}{the} precoding matrix in (\ref{18}), we get an optimal solution structure based on symbol-level extrapolation from the $n$-th symbol slot to $m$-th symbol slot, given by
\textcolor{my1}{
\begin{equation}
	\begin{aligned}
	&{\bf W}^{(m)} =\\&
	\frac{1}{K} \cdot {{\bf{H}}^H}{\left( {{\bf{H}}{{\bf{H}}^H}} \right)^{ - 1}}{\bf{U}}diag\left( \mathcal{F}_1\left(\ {\bf V}_{n}\right) \right)({{\bf A}\otimes {\bf I}_2 }){\bf s}_{\bf E}^{(n)}\left({{{\bf{\hat s}}}^{(m)}}\right)^{T},
	\end{aligned}
	\label{30}
\end{equation}}where the $\mathcal{F}_1\left(\ {\bf V}_{n}\right)\in \mathbb{R}^{2K\times1}$ represent the mapping function between ${\bf V}_n$ and ${{\bf \Omega}^{(m)}}$.\par 
Next, we will discuss how to calculate the symbol-scaling factors \textcolor{my}{${\bf \Omega}^{m}$} for the $m$-th symbol slot using the extrapolated matrix ${\bf V}_{m}$. When the problem $\mathcal{P}_3$ achieves the optimal solution for the $m$-th symbol slot, \textcolor{my}{the BS transmits at its maximum power and} we can obtain
\begin{equation}
({{\bf \Omega}^{(m)}})^T {\bf V}_{m}  {\bf \Omega}^{(m)} - p_0 = 0.
\label{31}
\end{equation}
Based on this formula, \textcolor{my}{we define a} power normalization factor described as
\begin{equation}
	f^{m}_{\mathrm{nor}}=\sqrt{({{\bf \Omega}^{(m)}})^T {\bf V}_{m}{\bf \Omega}^{(m)}}.
	\label{32}
\end{equation}
\textcolor{my}{We further introduce} ${\bf \Omega}^{(m)}_{\mathrm{fin}}\in \mathbb{R}^{2K\times1}$ \textcolor{my1}{to} represent the symbol-scaling factors that satisfy the power constraint \textcolor{my}{for the optimization problem}. The relationship between ${\bf \Omega}^{(m)}_{\mathrm{fin}}$ and ${\bf \Omega}^{(m)}$ can be given by
\begin{equation}
	{\bf \Omega}^{(m)}_{\mathrm{fin}} =\sqrt{p_{0}}\frac{{\bf \Omega}^{(m)}}{f^{m}_{\mathrm{nor}}}.
	\label{33}
\end{equation}
%ZF is a specific instance of SLP. We first analyze the ZF precoding from the viewpoint of symbol-scaling metric for the $m$-th symbol slot, establishing two fundamental criteria and extending them to more general scenarios. We initial the symbol-scaling factors ${\bf \Omega}^{\mathrm{ZF}} = {\bf 1}\in \mathbb{R}^{2K\times1}$.
%The magnitude of power normalization factor $n^{\mathrm{ZF}}_{\mathrm{nor}}$ is determined jointly by the channel and data symbols using the (\ref{27}), and the final symbol-scaling factors ${\bf \Omega}^{\mathrm{ZF}}_{\mathrm{fin}}$ can be obtained by (\ref{28}).\par
%
%
%In order to ensure that the performance of the obtained precoding matrix is better than ZF for SLP design, there are two criteria need to be satisfied: 1) before the power normalization, ${\bf \Omega}^{(m)}\succeq {\bf 1}$ for all data symbols, 2) after the power normalization, normalization factor should satisfy $n^{m}_{\mathrm{nor}}\leq n^{\mathrm{ZF}}_{\mathrm{nor}}$.\par 
\textcolor{my1}{Then we can} expand (\ref{31}) for the $m$-th symbol slot as
\begin{equation}
	\begin{aligned}
	&({\bf \Omega }^{(m)})^{T}{\bf V}_{m}{\bf \Omega}^{(m)} =\\
	&\sum_{j=1}^{2K} \sum_{i=1,i\neq j}^{2K}({{\bf V}_{m,i,i} ({\bf \Omega}^{(m)}_{i})^{2}}+{\bf V}_{m,i,j}{{\bf \Omega}^{(m)}_{i}}{{\bf \Omega}^{(m)}_{j}})+{p}_0,
	\end{aligned}
	 \label{34}
\end{equation}
where ${\bf \Omega}^{(m)}_{i}$ represents the $i$-th element \textcolor{my}{in} ${\bf \Omega}^{(m)}$. \textcolor{my}{(\ref{34}) is a series of equations presented in a matrix form, which can be observed as the sum} of $2K$ one-variable quadratic equations. Based on this \textcolor{my}{observation}, a closed-form iterative algorithm is proposed. 
\begin{algorithm}[h]
	\caption{SLE-CF-PSK Algorithm}
	\begin{algorithmic}
		\State ${\bf input:}$ ${\bf H}$, ${\mathcal{M}}$, ${\bf s}^{(n)}$, ${\bf s}^{(i)}$, ${M_{\mathrm{t}}}$, ${K}$, $R$, $N_{\mathrm{s}}$
		\State ${\bf output:}$ ${\bf W}_{i},i=1,2,\cdots,N_{\mathrm{s}}$
		\State ${\bf U} = {\bf I}_{M_{t}}\otimes[1,1]$;
%		\State Obtain ${\bf T}_{n} = diag(({\bf s}^{(n)})^{H}){\bf U}^{H}{({\bf H}{\bf H}^{H})}^{-1}{\bf U}diag({\bf s}^{(n)})$;
		\State Calculate ${\bf s}_{\bf E}^{(n)}$ using the (\ref{9}), (\ref{10}) and (\ref{21});
		\State Calculate ${\bf V}_{n}$ using (\ref{25});
		\For {$i = 1:N_{\mathrm{s}}$}
		\State Initial ${\bf \Omega}^{(i)} ={\bf 1}$, $n=0$;
		\State Obtain the ${\bf A}$ between the ${\bf s}^{(i)}$ and ${\bf s}^{(n)}$;
		\State Obtain the ${\bf T}_{i}$ and ${\bf V}_{i}$ using the (\ref{28}) and (\ref{29});
		\While {$n<R$}
		\State ${\bf q}^{(i)} = {\bf V}_{i}{\bf \Omega}^{(i)}$;
		\State $a=min({\bf q}^{(i)})$;
		\If {$a<0$}
		\State Find the $index$ of $a$ in the ${\bf q}^{(i)}$;
		\State Update the ${\bf \Omega}_{index}^{(i)}$ using (\ref{36});
		\State $n = n + 1$;
		\Else 
		\State break
		\EndIf 
		\EndWhile	
		\State Update ${\bf \Omega}_{\mathrm{fin}}^{(i)}$ using the (\ref{32}) and (\ref{33});
		\State \textcolor{my}{Obtain ${\bf W}^{(i)}$} using the (\ref{30});
		\EndFor	
	\end{algorithmic}
	\label{cf-psk}
\end{algorithm}
\par
\textcolor{my}{To be more specific, we} initialize the symbol-scaling factors ${\bf \Omega}^{(m)}={\bf 1}\in \mathbb{R}^{2K\times1}$. In order to approach the optimal solution as closely as possible, we update only one symbol-scaling factor in each iteration. \textcolor{my}{According to the (\ref{28}) and (\ref{29})}, We \textcolor{my}{introduce} ${\bf q}^{(m)}\in \mathbb{R}^{2K\times1}$ \textcolor{my}{that} can be expressed as
\textcolor{my}{
\begin{equation}
	{\bf q}^{(m)}=\Re \left(\left({{\bf A}\otimes {\bf I}_2 }\right)^{H}{\bf T}_{n}\left({{\bf A}\otimes {\bf I}_2 }\right)\right){\bf \Omega}^{(m)}.
	\label{35}
\end{equation}
}The \textcolor{my}{index of the} symbol-scaling factor is determined by the index of $min({\bf q})$. \textcolor{my}{In an iteration where we assume} that the $k$-th symbol-scaling factor for the $m$-th symbol slot is updated, the update formula is expressed as
\begin{equation}
	{\bf \Omega}^{(m)}_{k} = -\frac{\sum_{j\ne k,j=1}^{2K} {\bf V}_{m,k,j}{\bf \Omega}^{(m)}_{j}}{{\bf V}_{m,k,k}}.
	\label{36}
\end{equation}
When the algorithm reaches \textcolor{my}{its} specified iteration number or $min({\bf q}^{(m)}) > 0$, the iteration process terminates. It is important to note that the ${\bf q}^{(m)}$ will also be updated in each iteration. Finally, we normalize ${\bf \Omega}^{(m)}$ to satisfy the power constraint by using the (\ref{32}) and (\ref{33}) to obtain ${\bf \Omega}^{(m)}_{\mathrm{fin}}\in \mathbb{R}^{2K\times1}$.
\par
Within one transmission block, we assume that the \textcolor{my}{maximum} number of symbol slots is $N_{\mathrm{s}}$, the number of iterations is $R$, ${\bf s}^{(i)}\in \mathbb{C}^{K\times1}$ represents the data symbol vector for $i$-th symbol slot. \textcolor{my1}{The ${\bf \Omega}_{index}^{(i)}$ denotes that the $index$-th symbol-scaling factor is selected to update in the $i$-th symbol slot.} From the $n$-th symbol slot, we extrapolate the precoding matrices for all symbol slots. The closed-form iterative algorithm, based on symbol-level extrapolation, is presented in \textcolor{my}{Algorithm} \ref{cf-psk}.\par

The \textcolor{my}{proposed `SLE-CF-PSK'} iterative algorithm updates only one symbol-scaling factor at each iteration, and is able to extrapolate a near-optimal performance, which will be shown numerically in \textcolor{my1}{Section VI.} \textcolor{my}{To further reduce the complexity, we design} a sub-optimal closed-form solution based on SLE to update \textcolor{my}{all} symbol-scaling factors \textcolor{my}{simultaneously}. \textcolor{my}{The sub-optimal algorithm begins by initializing} the symbol-scaling factors ${\bf \Omega}^{(m)}={\bf 1}\in \mathbb{R}^{2K\times1}$. \textcolor{my}{Then} the symbol-scaling factors ${\bf \Omega}^{(m)}_{\mathrm{sub}}\in \mathbb{R}^{2K\times1}$ for the $m$-th symbol slot \textcolor{my}{are updated based on the following principle:}
\begin{equation}
 {\bf \Omega}^{(m)}_{\mathrm{sub}}=max\left(-\frac{{\bf V}_{m}{\bf \Omega}^{(m)}-diag({\bf V}_{m})}{2diag({\bf V}_{m})},1\right) 
 \label{37}
\end{equation}
Then, we get the final magnitude of the symbol-scaling factor ${\bf \Omega}^{(m)}_{\mathrm{sub,fin}}\in \mathbb{R}^{2K\times1}$ by substituting ${\bf \Omega}^{(m)}_{\mathrm{sub}}$ into the (\ref{27}) and (\ref{28}). The closed-form sub-optimal algorithm \textcolor{my}{is presented in Algorithm} \ref{sub-psk}.\par

\begin{algorithm}%[!b]
	\caption{SLE-Sub-PSK Algorithm}
	\begin{algorithmic}
		\State ${\bf input:}$ ${\bf H}$, ${\mathcal{M}}$, ${\bf s}^{(n)}$, ${\bf s}^{(i)}$, ${M_{\mathrm{t}}}$, ${K}$, $R$, $N_{\mathrm{s}}$
		\State ${\bf output:}$ ${\bf W}_{i},i=1,2,\cdots,N_{\mathrm{s}}$
		\State ${\bf U} = {\bf I}_{M_{t}}\otimes[1,1]$;
%		\State Obtain ${\bf T}_{n} = diag(({\bf s}^{(n)})^{H}){\bf U}^{H}{({\bf H}{\bf H}^{H})}^{-1}{\bf U}diag({\bf s}^{(n)})$;
		\State Calculate ${\bf s}_{\bf E}^{(n)}$ using the (\ref{9}), (\ref{10}) and (\ref{21});
		\State Calculate ${\bf V}_{n}$ using (\ref{25});
		\For {$i = 1:N_{\mathrm{s}}$}
		\State Initial ${\bf \Omega}^{(i)} ={\bf 1}$;
		\State Obtain the ${\bf A}$ between the ${\bf s}^{(i)}$ and ${\bf s}^{(n)}$;
		\State Obtain the ${\bf T}_{i}$ and ${\bf V}_{i}$ using the (\ref{28}) and (\ref{29});
	    \State Obtain the ${\bf \Omega}^{(i)}_{\mathrm{sub}}$	
		\State Update ${\bf \Omega}^{(i)}_{\mathrm{sub,fin}}$ using the (\ref{32}) and (\ref{33});
		\State Obtain ${\bf W}^{(i)}$ using the (\ref{30});
		\EndFor	
	\end{algorithmic}
\label{sub-psk}
\end{algorithm}

\subsection{The case of $\mathcal{M}$-QAM}
\textcolor{my1}{Based on the symbol-scaling metric in \cite{li2020interference}}, the optimization problem that maximizes the CI effect \textcolor{my1}{using QAM modulation for} the $n$-th symbol slot is give by
\textcolor{my1}{
\begin{equation}
	\begin{aligned}
		&\mathcal{P}_4: {\kern 3pt} \mathop {\max }\limits_{{{\bf{W}}_{n}}, {\kern 1pt} t} {\kern 3pt} t \\
		&{\kern 0pt} s. t. {\kern 10pt} {{\bf{h}}_k^T}{{\bf W}_{n}{\bf s}^{(n)}} = \alpha _{k,\mathcal{A}}^{(n)}s_{k,\mathcal{A}}^{(n)}+\alpha _{k,\mathcal{B}}^{(n)}s_{k,\mathcal{B}}^{(n)}, {\kern 3pt} \forall k \in {\cal K}, \\
		&{\kern 24pt} t \le \alpha_i^{\cal O}, {\kern 3pt} \forall \alpha_i^{\cal O} \in {\cal O}^{(n)},\\
		&{\kern 24pt} t = \alpha_j^{\cal I}, {\kern 3pt} \forall \alpha_j^{\cal I} \in {\cal I}^{(n)},\\
		&{\kern 22pt} \left\| {{\bf{W}}_{n}{\bf s}^{(n)}} \right\|_2^2 \le {p_0},
		\label{eq_13}
	\end{aligned}
\end{equation}}where $\mathcal{K}=\left\{ {1,2,\cdots,K}\right\}$, $\alpha_i^{\cal O}$ represents the $i$-th element in set ${\cal O}^{(n)}$,  $\alpha_j^{\cal I}$ represents the $j$-th element in set ${\cal I}^{(n)}$, $p_{0}$ is the total power at the transmitter. 
%\textcolor{my1}{For QAM modulation,} $\mathcal{O}^{(n)}$ and $\mathcal{I}^{(n)}$ for the $n$-th symbol slot can be expressed as\par
%\begin{equation}
%	{\cal O}^{(n)} \cup {\cal I}^{(n)} = \left\{ {\alpha _{1,\cal A}^{(n)} ,{\alpha _{1,\cal B}^{(n)}},{\alpha _{2,\cal A}^{(n)}},{\alpha _{2,\cal B}^{(n)}}, \cdots ,{\alpha _{K,\cal A}^{(n)}},{\alpha _{K,\cal B}^{(n)}}} \right\},
%	\label{38}
%\end{equation}
%and 
%\begin{equation}
%	card\left| {\cal O}^{(n)} \right| + card\left| {\cal I}^{(n)} \right| = 2K,
%	\label{39}
%\end{equation}
\textcolor{my1}{It is easy to verify that} $\mathcal{P}_4$ \textcolor{my}{is convex and} can be directly solved via convex optimization tools. In \cite{li2020interference}, the authors derive the optimal solution structure of the precoding matrix. For the $n$-th symbol slot, the precoding matrix can be expressed as
\begin{equation}
	\begin{aligned}
		{\bf W}^{(n)} &=\\&
		\frac{1}{K} \cdot {{\bf{H}}^H}{\left( {{\bf{H}}{{\bf{H}}^H}} \right)^{ - 1}}{\bf{U}}diag\left( {\bf \Omega}^{(n)}\right){\bf s}_{\bf E}^{(n)}\left({{{\bf{\hat s}}}^{(n)}}\right)^{T}.
	\end{aligned}
	\label{41}
\end{equation}
where ${{\bf{\hat s}}^{(n)}}$, ${\bf U}$ and ${\bf s}_{\bf E}^{(n)}$ can be obtained by using the (\ref{19}), (\ref{20}) and (\ref{21}). Therefore, how to efficiently compute the symbol-scaling factors with low complexity is important for CI-SLP design. 
\par
When we obtain ${\bf \Omega}^{(n)}$, the optimal precoding matrix \textcolor{my}{can be directly obtained} by using the (\ref{41}). Therefore, the key issue is how to efficiently compute the symbol-scaling factors with low complexity. In \cite{li2020interference}, the authors utilize the Lagrangian dual and \textcolor{my1}{KKT} conditions to transform the original problem $\mathcal{P}_3$ into \textcolor{my}{simpler} forms. \textcolor{my}{More specifically, the original} optimization problem can be simplified \textcolor{my}{into}

%In \cite{li2020interference}, the authors utilize the Lagrangian dual and Karush-Kuhn-Tucker (KKT) conditions to transform the original problem $\mathcal{P}_4$ into a more simple forms. The optimization problem for the $n$-th symbol slot can be simplified as
\textcolor{my1}{
\begin{equation}
	\begin{aligned}
		&\mathcal{P}_5: {\kern 3pt} \mathop {\min }\limits_{{{\bf \Omega}^{(n)}}, {\kern 1pt} t} {\kern 3pt} -t \\
		&{\kern 0pt} s. t. {\kern 10pt} ({{\bf \Omega}^{(n)}})^T {\bf V}_{n}{\bf \Omega}^{(n)} - p_0 = 0, \\
		&{\kern 24pt} t - {\bf \Omega}_{i}^{(n)} \le 0, {\kern 3pt} \forall i\in {\cal K},\\
		&{\kern 24pt} t - \alpha_i^{\cal O} \le 0, {\kern 3pt} \forall \alpha_i^{\cal O} \in {\cal O}^{(n)},\\
		&{\kern 24pt} t - \alpha_j^{\cal I} = 0 , {\kern 3pt} \forall \alpha_j^{\cal I} \in {\cal I}^{(n)},
		\label{42}
	\end{aligned}
\end{equation}}where the ${\bf V}_{n}$ can be obtained by using the (\ref{25}). \textcolor{my}{The optimization $\mathcal{P}_6$ on ${\bf \Omega}^{(m)}$} for the $m$-th symbol slot \textcolor{my}{is in the same form as $\mathcal{P}_5$ above.}\par

Within one transmission block, we exploit the relationships between \textcolor{my}{the} $n$-th symbol slot and \textcolor{my}{the} $m$-th symbol slot to connect $\mathcal{P}_5$ and $\mathcal{P}_6$. \textcolor{my}{We} employ ${\bf Q}_n$ from the $n$-th symbol slot to extrapolate ${\bf V}_{m}$, thereby establishing \textcolor{my}{the} connection \textcolor{my}{of the precoding matrix} between the two \textcolor{my}{symbol slots} to reduce the complexity. The relationship between ${\bf Q}_n$ and ${\bf V}_{m}$ can be expressed as
\begin{equation}
	{\bf V}_{m} = \Re\left({\bf T}_{m}\right),
	\label{45}
\end{equation}
\begin{equation}
	{\bf T}_{m} = {diag\left( ({\bf s}_{\bf E}^{(m)} )^{H} \right) {\bf Q}_n diag\left( {{\bf s}_{\bf E}^{(m)}}\right)}.
	\label{44}
\end{equation}
Based on the above derivations and the specific solution structure of precoding matrix in (\ref{41}), we get an optimal solution structure based on symbol-level extrapolation from the $n$-th symbol slot to $m$-th symbol slot, given by
\textcolor{my}{
	\begin{equation}
		\begin{aligned}
			{\bf W}^{(m)} &=\\&
			\frac{1}{K} \cdot {{\bf{H}}^H}{\left( {{\bf{H}}{{\bf{H}}^H}} \right)^{ - 1}}{\bf{U}}diag\left( \mathcal{F}_2\left(\ {\bf Q}_n\right) \right){\bf s}_{\bf E}^{(m)}\left({{{\bf{\hat s}}}^{(m)}}\right)^{T},
		\end{aligned}
		\label{444}
	\end{equation}
where the $\mathcal{F}_2\left(\ {\bf Q}_n\right)\in \mathbb{R}^{2K\times1}$ represent the mapping function between ${\bf Q}_n$ and ${{\bf \Omega}^{(m)}}$.}\par

Next, we will describe how to utilize the matrix ${\bf V}_{m}$ to determine the symbol-scaling factors ${\bf \Omega}^{(m)}$. QAM modulation can be considered as an extension in terms of the PSK modulation. Therefore, we can extend the SLE-CF-PSK algorithm to QAM modulation considering the case where the set ${\cal I}^{(m)}$ is not empty. We assume that the ${\cal D}^{(m)}\in \mathbb{R}^{card\left|{\cal I}^{(m)}\right| \times 1}$ is the index set corresponds to the set ${\cal I}^{(m)}$, and the ${\cal L}^{(m)}\in \mathbb{R}^{card\left|{\cal O}^{(m)}\right| \times 1}$ is the index set corresponds to the set ${\cal O}^{(m)}$.

\textcolor{my}{To be more specific, we} initialize the symbol-scaling factors ${\bf \Omega}^{(m)}={\bf 1}\in \mathbb{R}^{2K\times1}$. In order to approach the optimal solution as closely as possible, we update only one symbol-scaling factor in each iteration. \textcolor{my}{According to the (\ref{45}) and (\ref{44})}, We \textcolor{my}{introduce} ${\bf q}^{(m)}\in \mathbb{R}^{2K\times1}$ and ${\bf q}^{(m)}_{\cal L}\in \mathbb{R}^{card\left|{\cal O}^{(m)}\right| \times 1}$ \textcolor{my}{that} can be expressed as

\begin{equation}
	{\bf q}^{(m)}=\Re\left\{{diag\left( ({\bf s}_{\bf E}^{(m)} )^{H} \right) {\bf Q}_n diag\left( {{\bf s}_{\bf E}^{(m)}}\right)}\right\}{\bf \Omega}^{(m)},
	\label{46}
\end{equation}
and
\begin{equation}
	{\bf q}^{(m)}_{\cal L}={\bf q}^{(m)}\left({\cal L}^{(m)}\right) \in \mathbb{R}^{card\left|{\cal O}^{(m)}\right| \times 1}.
	\label{47}
\end{equation}
The \textcolor{my}{index of the} symbol-scaling factor is determined by the index of $min({\bf q}^{(m)}_{\cal L})$. \textcolor{my}{In an iteration where we assume} that the $k$-th symbol-scaling factor for the $m$-th symbol slot is updated, the update formula is expressed as
\begin{equation}
	{\bf \Omega}^{(m)}_{k} = -\frac{\sum_{j\ne k,j=1}^{2K} {\bf V}_{m,k,j}{\bf \Omega}^{(m)}_{j}}{{\bf V}_{m,k,k}}.
	\label{48}
\end{equation}
When the iteration reaches its iteration number or $min({\bf q}_{\cal L}^{(m)}) > 0$, the iteration process terminates.
Finally, we normalize ${\bf \Omega}^{(m)}$ to satisfy the power constraint by using the (\ref{32}) and (\ref{33}) to obtain the ${\bf \Omega}^{(m)}_{\mathrm{fin}}\in \mathbb{R}^{2K\times1}$. 
\par

Within one transmission block, we assume that the \textcolor{my}{maximum} number of symbol slots is $N_{\mathrm{s}}$, the number of iterations is $R$, ${\bf s}^{(i)}\in \mathbb{C}^{K\times1}$ represents the data symbol vector for $i$-th symbol slot. \textcolor{my1}{The ${\bf \Omega}_{index}^{(i)}$ denotes that the $index$-th symbol-scaling factor is selected to update in the $i$-th symbol slot.} From the $n$-th symbol slot, we extrapolate the precoding matrices for all symbol slots. The closed-form iterative algorithm, based on symbol-level extrapolation, is presented in \textcolor{my}{Algorithm} \ref{cf-qam}.
\par

\begin{algorithm}%[!b]
	\caption{SLE-CF-QAM Algorithm}
	\begin{algorithmic}
		\State ${\bf input:}$ ${\bf H}$, ${\bf s}^{(n)}$, ${\bf s}^{(i)}$, ${M_{\mathrm{t}}}$, ${K}$, $R$, $N_{\mathrm{s}}$
		\State ${\bf output:}$ ${\bf W}_{i},i=1,2,\cdots,N_{\mathrm{s}}$
		\State ${\bf U} = {\bf I}_{M_{t}}\otimes[1,1]$;
%		\State Obtain ${\bf T}_{n} = diag(({\bf s}^{(n)})^{H}){\bf U}^{H}{({\bf H}{\bf H}^{H})}^{-1}{\bf U}diag({\bf s}^{(n)})$;
		\State Calculate ${\bf s}_{\bf E}^{(n)}$ using the (\ref{13}) and (\ref{21});
		\State Calculate ${\bf V}_{n}$ using the (\ref{25});
		\For {$i = 1:N_{\mathrm{s}}$}
		\State Initial ${\bf \Omega}^{(i)} ={\bf 1}$, $n=0$;
		\State Obtain the ${\bf A}$ between the ${\bf s}^{(i)}$ and ${\bf s}^{(n)}$;
		\State Obtain the ${\bf T}_{i}$ and ${\bf V}_{i}$ using (\ref{45}) and (\ref{44});
		\If {${\bf{q}}^{(m)}_{\mathcal{L}}\ne \emptyset$}
		\While {$n<R$}
		\State ${\bf q}^{(m)} = {\bf V}_{i}{\bf \Omega}^{(i)}$;
		\State Obtain ${\bf{q}}^{(i)}_{\mathcal{L}}$ using ({\ref{47}})
		\State $a=min({\bf{q}}^{(i)}_{\mathcal{L}})$;
		\If {$a<0$}
		\State Find the $index$ of $a$ in the ${\bf{q}}^{(m)}_{\mathcal{L}}$;
		\State Update the ${\bf \Omega}_{index}^{(i)}$ using (\ref{48});
		\State $n = n + 1$;
		\Else 
		\State break
		\EndIf 
		\EndWhile
		\EndIf	
		\State Update ${\bf \Omega}^{(i)}_{\mathrm{fin}}$ using the (\ref{32}) and (\ref{33});
		\State Obtain ${\bf W}^{(i)}$ using the (\ref{444});
		\EndFor	
	\end{algorithmic}
\label{cf-qam}
\end{algorithm}

The \textcolor{my}{proposed `SLE-CF-QAM'} iterative algorithm updates only one symbol-scaling factor at each iteration, and is able to extrapolate a near-optimal performance, which will be shown numerically in \textcolor{my1}{Section VI.} \textcolor{my}{To further reduce the complexity, we design} a sub-optimal closed-form solution based on SLE to update \textcolor{my}{all} symbol-scaling factors \textcolor{my}{simultaneously}. \textcolor{my}{The sub-optimal algorithm begins by initializing} the symbol-scaling factors ${\bf \Omega}^{(m)}={\bf 1}\in \mathbb{R}^{2K\times1}$. \textcolor{my}{Then} the symbol-scaling factors ${\bf \Omega}^{(m)}_{\mathrm{sub}}\in \mathbb{R}^{2K\times1}$ for the $m$-th symbol slot \textcolor{my}{are updated based on the following principle:}
\begin{equation}
	{\bf \Omega}^{(m)}_{\mathrm{sub}}=max\left(-\frac{{\bf V}_{m}{\bf \Omega}^{(m)}-diag({\bf V}_{m})}{2diag({\bf V}_{m})},1\right), 
	\label{49}
\end{equation}
and
\begin{equation}
	{\bf \Omega}^{(m)}_{\mathrm{sub}}\left(\mathcal{D}^{(m)}\right) = {\bf 1}. 
	\label{50}
\end{equation}
Then, we get the final magnitude of the symbol-scaling factors ${\bf \Omega}^{(m)}_{\mathrm{sub},\mathrm{fin}}\in \mathbb{R}^{2K\times1}$ by substituting ${\bf \Omega}^{(m)}_{\mathrm{sub}}$ into the (\ref{32}) and (\ref{33}). The closed-form sub-optimal algorithm is presented in Algorithm \ref{sub-qam}.\par

\begin{algorithm}%[!b]
	\caption{SLE-Sub-QAM Algorithm}
	\begin{algorithmic}
		\State ${\bf input:}$ ${\bf H}$, ${\bf s}^{(n)}$, ${\bf s}^{(i)}$, ${M_{\mathrm{t}}}$, ${K}$, $R$, $N_{\mathrm{s}}$
		\State ${\bf output:}$ ${\bf W}_{i},i=1,2,\cdots,N_{\mathrm{s}}$
		\State ${\bf U} = {\bf I}_{M_{t}}\otimes[1,1]$;
%		\State Obtain ${\bf T}_{n} = diag(({\bf s}^{(n)})^{H}){\bf U}^{H}{({\bf H}{\bf H}^{H})}^{-1}{\bf U}diag({\bf s}^{(n)})$;
		\State Calculate ${\bf s}_{\bf E}^{(n)}$ using the (\ref{13}) and (\ref{21});
		\State Calculate ${\bf V}_{n}$ using the (\ref{25});
		\For {$i = 1:N_{\mathrm{s}}$}
		\State Initial ${\bf \Omega}^{(i)} ={\bf 1}$;
		\State Obtain the ${\bf A}$ between the ${\bf s}^{(i)}$ and ${\bf s}^{(n)}$;
		\State Obtain the ${\bf T}_{i}$ and ${\bf V}_{i}$ using (\ref{45}) and (\ref{44});
		\State Obtain the ${\bf \Omega}^{(i)}_{\mathrm{sub}}$ using the (\ref{49}) and (\ref{50});	
		\State Update ${\bf \Omega}^{(i)}_{\mathrm{sub,fin}}$ using the (\ref{32}) and (\ref{33});	
		\State Obtain ${\bf W}^{(i)}$ using the (\ref{444});
		\EndFor	
	\end{algorithmic}
\label{sub-qam}
\end{algorithm}

\section{Deep Unfolding for SLE-based Iterative algorithm}

To reduce complexity while maintaining the performance, we employ the deep unfolding to expand the iterative algorithm \textcolor{my}{and construct the SLE-Net}. This approach leverages the parallel \textcolor{my}{processing capability} of \textcolor{my}{GPUs} to decrease time complexity. We introduce a deep unfolding framework which can be applicable \textcolor{my}{to both} PSK and QAM modulation. \textcolor{my}{Essentially}, QAM modulation can be considered \textcolor{my}{as} an extension of PSK modulation for the network. Therefore, \textcolor{my}{below} we choose the QAM modulation as \textcolor{my}{the example to describe our network.}\par
\begin{figure*}[h]
	\centering
	\includegraphics[width=1\textwidth]{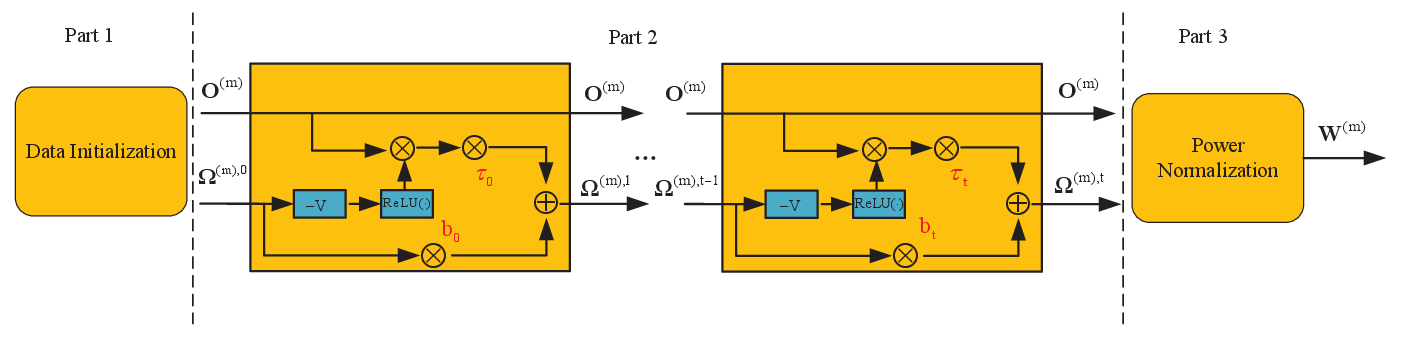}
	\captionsetup{labelformat=default,labelsep=space}
	\caption{\textcolor{my1}{the diagram of the SLE-Net}}
	\label{network}
\end{figure*}
\subsection{\textcolor{my}{Network Architecture}}
The overall framework of the network is illustrated in Fig. \ref{network} and can be divided into three parts. In the first part, data initialization is performed. Given the CSI and the data symbol vectors, we \textcolor{my1}{aim to extrapolate the matrices ${\bf V}$ and get the matrix ${\bf O}\in \mathbb{R}^{2K\times N_{\mathrm{s}}}$, which is an indicator matrix that shows whether symbol-scaling factors can be scaled for all the symbol slots.} For convenience of description, we illustrate the \textcolor{my}{extrapolation process for} the $m$-th symbol slot as an example. ${\bf \Omega}^{(m),0}={\bf 1}\in \mathbb{R}^{2K\times1}$ represent the initial symbol-scaling factors. \textcolor{my}{Based on the definition,} ${\bf O}^{(m)}\in \mathbb{R}^{2K\times1}$ represents \textcolor{my1}{$m$-th vector in matrix ${\bf O}$} to indicate whether symbol-scaling factors can be scaled for the $m$-th symbol slot, \textcolor{my}{whose} the expression is \textcolor{my}{given by}

\begin{equation}
	{\bf O}_{i}^{(m)} =  \left\{\begin{array}{ll}
		1, i\in \mathcal{O}^{(m)},\\
		0, i\in \mathcal{I}^{(m)}.
	\end{array}\right.
\label{51}
\end{equation}

The second part consists of iterative unfolding layers, where the number of layers in the network corresponds to the iteration numbers of the SLE-CF-QAM algorithm. Specifically, the following updates are used for \textcolor{my}{$r = 1,2,\cdots,R$:}
\begin{equation}
	{\bf \Omega}^{(m),r} = {\bf g}(-{\bf V}{\bf \Omega}^{(m),r-1})\odot{\bf O}^{(m)} \odot {\mathbf{\tau }}_{r-1} +b_{r-1}{\bf \Omega}^{(m),r-1}, 
\end{equation}
where ${\bf \Omega}^{(m),r}\in \mathbb{R}^{2K\times1}$ represents the symbol-scaling factors after the $r$-th iteration, $b_{r}\in \mathbb{R}$ and ${\mathbf{\tau }}_{r}\in \mathbb{R}^{2K\times1}$ represents the learning parameters in the $r$-th iterative unfolding layer, ${\bf g}$ is the activation function and can be expressed as
\begin{equation}
	{\bf g}({\bf x}) = max({\bf x},{\bf 0}).
\end{equation}
\par
The third part involves normalizing the symbol-scaling factors \textcolor{my}{to meet the} power constraint \textcolor{my}{at the BS}, and then substituting the symbol-scaling factors into the closed-form solution to obtain the final precoding matrix.
\par
\subsection{\textcolor{my}{Training the Network}}
Training deep networks is a difficult task due to \textcolor{my1}{the issues of} vanishing gradients, saturation of the activation functions, sensitivity to initialization\textcolor{my}{\cite{roodschild2020new}}. Based on the simplified original optimization problem $\mathcal{P}_6$, we \textcolor{my}{design} an unsupervised loss function to maximize the CI effects by staying away from the detection threshold, transforming the network from data-driven to model-driven, \textcolor{my}{which helps enhance} the interpretability of the network. We assume that the output of the network is \textcolor{my}{${\bf \hat{y}}_{n}\in\mathbb{R}^{2K\times1}$ for the $n$-th sample}, the loss function can be expressed as
\begin{equation}
	loss =-\frac{1}{N}\sum_{n=1}^{N} min({\bf \hat{y}}_{n}),
\end{equation}
where $N$ is the batchsize, ${\bf\hat{y}}_{n}$ denotes the output of the $n$-th training sample. The loss function is minimized by \textcolor{my1}{the adaptive moment estimation (ADAM)} algorithm in \cite{kingma2014adam} until the network converges. \textcolor{my}{The SLE-Net algorithm is presented in Algorithm \ref{net}.}

\begin{algorithm}%[!b]
	\caption{SLE-Net Algorithm}
	\begin{algorithmic}
		\State ${\bf input:}$ ${\bf H}$, ${\bf s}^{(n)}$, ${\bf s}^{(i)}$, ${M_{\mathrm{t}}}$, ${K}$, $R$, $N_{\mathrm{s}}$, epoch numbers $E$, learning rate $\gamma$, batchsize $N$
		\State ${\bf output:}$ ${\bf W}_{i},i=1,2,\cdots,N_{\mathrm{s}}$
		\For {$e = 1:E$}
		\State Update the parameters (learning rate $\gamma$) by using the ADAM algorithm to minimize the loss;
		\EndFor
		\State ${\bf U} = {\bf I}_{M_{t}}\otimes[1,1]$;
%		\State Obtain ${\bf T}_{n} = diag(({\bf s}^{(n)})^{H}){\bf U}^{H}{({\bf H}{\bf H}^{H})}^{-1}{\bf U}diag({\bf s}^{(n)})$;
		\If {$\mathcal{M}$-QAM} 
		\State Calculate ${\bf s}_{\bf E}^{(n)}$ using the (\ref{13}) and (\ref{21});
		\Else
		\State Calculate ${\bf s}_{\bf E}^{(n)}$ using the (\ref{9}), (\ref{10}) and (\ref{21});
		\EndIf
		\State Calculate ${\bf V}_{n}$ using the (\ref{25});
		\For {$i = 1:N_{\mathrm{s}}$}
		\State Initial ${\bf \Omega}^{(i)} ={\bf 1}$, $n=0$;
		\State Obtain the ${\bf O}^{(i)}$ using the (\ref{51});
		\State Obtain the ${\bf A}$ between the ${\bf s}^{(i)}$ and ${\bf s}^{(n)}$;
		\If {$\mathcal{M}$-QAM} 
		\State Obtain the ${\bf T}_{i}$ and ${\bf V}_{i}$ using (\ref{45}) and (\ref{44});
		\Else
		\State Obtain the ${\bf T}_{i}$ and ${\bf V}_{i}$ using (\ref{28}) and (\ref{29});
		\EndIf
		\State Obtain the ${\bf \Omega}^{(i)}_{\mathrm{net}}$;
		\If {$\mathcal{M}$-QAM} 	
		\State Obtain ${\bf W}^{(i)}$ by using (\ref{444});
		\Else
		\State Obtain ${\bf W}^{(i)}$ by using (\ref{30});
		\EndIf
		\EndFor	
	\end{algorithmic}
\label{net}
\end{algorithm}

\section{Complexity Analysis}

In this section, we evaluate the complexity of the proposed algorithm in terms of the required number of multiplications and additions. For the purpose of simplifying the analysis, we \textcolor{my}{evaluate} the complexity \textcolor{my}{in terms of the required} complex number multiplication and addition operations. In the application scenario of this paper, we assume that the number of antennas $M_{\mathrm{t}}$ at the transmitter is equal to the number of users $K$. Within one transmission block, we assume that the number of symbol slots for transmission is $N_{\mathrm{s}}$, and the number of iterations is $R$. We have proposed algorithms for both PSK and QAM modulation schemes. Therefore, we first analyze the complexity of the PSK algorithm and subsequently extend the results to QAM algorithm.\par
Based on the rules of matrix inversion and multiplication operations, the complexity of ${\bf H}^{H}{({\bf H}{\bf H}^{H})}^{-1}$ is $\mathcal{O}(5K^3-K^2+K)$. The complexity of ZF precoding for $N_{\mathrm{s}}$ symbol slots is given by
\begin{equation}
	\mathcal{O}_{\mathrm{ZF}}=(5K^3+2K^2)N_{\mathrm{s}}.
\end{equation}

\par 

\subsection{Complexity of SLE-Sub-PSK and SLE-CF-PSK}

Whether it is an iterative algorithm or a sub-optimal closed-form algorithm for PSK modulation in this paper, the overall complexity of the algorithm can be divided into three parts within a transmission block. The first part involves the extrapolation of the matrix ${\bf V}$ and the vector ${\bf s}_{\bf E}$ for all symbol slots. The second part is to calculate the symbol-scaling factors based on the matrix ${\bf V}$. The third part is the normalization of the obtained symbol-scaling factors, substituting symbol-scaling factors into the (\ref{30}) to get the precoding matrix.
\par

According to the formula (\ref{27}), (\ref{28}) and (\ref{29}), the complexity to extrapolate the matrix ${\bf V}$ and the vector ${\bf s}_{\bf E}$ for $N_{\mathrm{s}}$ symbol slots is given by
\begin{equation}
\mathcal{O}_{\mathrm{e}}=3K^3+(8N_{\mathrm{s}}+4)K^2+(5N_{\mathrm{s}}+3)K.
\end{equation}
For the sub-optimal SLE-Sub-PSK algorithm, the complexity of the formula (\ref{37}) to calculate the symbol-scaling factors for $N_{\mathrm{s}}$ symbol slots is given by
\begin{equation}
	\mathcal{O}_{\mathrm{sub}}=(8K^2+4K)N_{\mathrm{s}}.
\end{equation}
For the SLE-CF-PSK algorithm, the complexity to calculate the symbol-scaling factors through $R$ iterations for $N_{\mathrm{s}}$ symbol slots is given by
\begin{equation}
	\mathcal{O}_{\mathrm{i}}=(8K^2+5K+1)RN_{\mathrm{s}}.
\end{equation}
Next, the complexity of the third part to normalize the symbol-scaling factors and substitute them into the formula (\ref{30}) to calculate the precoding matrix for $N_{\mathrm{s}}$ symbol slots is given by 
\begin{equation}
	\mathcal{O}_{\mathrm{n}}=(9K^3+13K^2+7K-1)N_{\mathrm{s}}.
\end{equation}  
\par
Based on the above analysis, we can obtain the overall complexity of the SLE-Sub-PSK and SLE-CF-PSK within one transmission block as follows

\begin{equation}
	\begin{aligned}
		\mathcal{O}_{\mathrm{sub}}^{\mathrm{psk}}&  =\mathcal{O}_{\mathrm{e}}+\mathcal{O}_{\mathrm{sub}}+\mathcal{O}_{\mathrm{n}}\\
		&=(9N_{\mathrm{s}}+3)K^3+(29N_{\mathrm{s}}+4)K^2+(16N_{\mathrm{s}}+3)K-N_{\mathrm{s}},
	\end{aligned}
\end{equation}

\begin{equation}
	\begin{aligned}
		\mathcal{O}_{\mathrm{ite}}^{\mathrm{psk}} &=\mathcal{O}_{\mathrm{e}}+\mathcal{O}_{\mathrm{i}}+\mathcal{O}_{\mathrm{n}}\\
		&=(9N_{\mathrm{s}}+3)K^3+(21N_{\mathrm{s}}+8RN_{\mathrm{s}}+4)K^2\\
		&+(12N_{\mathrm{s}}+5RN_{\mathrm{s}}+3)K+(R-1)N_{\mathrm{s}}.
	\end{aligned}
\end{equation}
\par

The complexity of optimization problems $\mathcal{P}_1$ when using interior-point methods to solve is influenced by the requirement of the accuracy $\varepsilon$ and the number of constraints. According to \cite{li2018interference}, The complexity of the $\mathcal{P}_1$ can be approximately expressed as 

\begin{equation}
	\begin{aligned}
	\mathcal{O}_{\mathrm{opt}} &= [28K^3 +49K^2+4K \\
	&+\ln(\frac{2K^2+12K+2}{\varepsilon})(K+2)^{0.5}(2K^3 + 3K^3 + 2K)]N_{\mathrm{s}}.
	\end{aligned} 
\end{equation}

\par
The complexity of the conventional algorithm in \cite{li2018interference}, denoted as the CI-CF-PSK, can be given by
\begin{equation}
	\mathcal{O}_{\mathrm{ci}} = (28K^3 +49K^2+4K + \sum_{r=0}^{R}(2r^2+3r+1))N_{\mathrm{s}}. 
\end{equation}

\par

\subsection{Complexity of SLE-Net}

We assume that the number of iteration unfolding layers is equal to the number of iterations $R$. As described earlier, it is known that the network consists of three parts. For the SLE-Net algorithm, the complexity of the first part and the third part are approximately equal to the $\mathcal{O}_{\mathrm{e}}$ and $\mathcal{O}_{\mathrm{n}}$. The complexity of the iteration unfolding layers for $N_{\mathrm{s}}$ symbol slots is given by
\begin{equation}
	\mathcal{O}_{\mathrm{d}}=(8K^2+6K)RN_{\mathrm{s}}.
\end{equation}
Therefore, for the PSK modulation, the complexity of the SLE-Net can be given by
\begin{equation}
	\begin{aligned}
		\mathcal{O}_{\mathrm{net}} &=\mathcal{O}_{\mathrm{e}}+\mathcal{O}_{\mathrm{d}}+\mathcal{O}_{\mathrm{n}}\\
		&=(9N_{\mathrm{s}}+3)K^3+(21N_{\mathrm{s}}+8RN_{\mathrm{s}}+4)K^2\\
		&+(12N_{\mathrm{s}} + 6RN_{\mathrm{s}}+3)K - N_{\mathrm{s}}.
	\end{aligned}
\end{equation}

\par
Compared to the proposed algorithm for PSK modulation, the algorithm for QAM modulation only increase the complexity of determining the set $\mathcal{O}$ and $\mathcal{I}$ for all the symbol slots. The complexity of this calculation can be negligible in comparison to the entire algorithm. Therefore, the complexities of the proposed algorithm for PSK modulation and QAM modulation is nearly identical. We assume that the $M_{\mathrm{t}}=K=8$, $R=5$, $N_{\mathrm{s}}=8000$, $\varepsilon=0.001$, and provide the number of the floating operations of different precoding schemes for PSK modulation in TABLE \ref{tab}.

\begin{table*}[ht]
	\centering
	\renewcommand{\arraystretch}{1.5} % 设置行间距的缩放因子 
	\caption{The number of FLOPS}
	\label{tab}
	\begin{tabular*}{\linewidth}{@{\extracolsep{\fill}}cccccc@{}}
		\hline\specialrule{0.1em}{0pt}{0.5pt}
		\hspace{2em} &$4\times4$ & $8\times8$& $12\times12$&$16\times16$ &$20\times20$\hspace{2em}\\ 
		\hline
		\hspace{2em}ZF &$2.816\times10^{6}$ &$2.1504\times10^{7}$ &$7.1424\times10^{7}$ &$1.67936\times10^{8}$ &$3.264\times10^{8}$ \hspace{2em}\\
		\hline
		\hspace{2em}CI-OPT &$1.082591\times10^{8}$ &$1.0935\times10^{9}$ &$4.407524\times10^{9}$ &$1.199742\times10^{10}$ &$2.622763\times10^{10}$\hspace{2em}\\
		\hline
		\hspace{2em}CI-CF-PSK &$2.2024\times10^{7}$ &$1.4132\times10^{8}$ &$4.45192\times10^{8}$ &$1.019656\times10^{9}$ &$1.950728\times10^{9}$ \hspace{2em}\\
		\hline
		\hspace{2em}SLE-CF-PSK &$1.3568268\times10^{7}$ &$7.043382\times10^{7}$ &$1.982138\times10^{8}$ &$4.245574\times10^{8}$ &$7.771137\times10^{8}$\hspace{2em}\\
		\hline
		\hspace{2em}SLE-Sub-PSK &$8.824268\times10^{6}$ &$5.272982\times10^{7}$ &$1.593578\times10^{8}$ &$3.563574\times10^{8}$ &$6.713777\times10^{8}$ \hspace{2em}\\
		\hline
		\hspace{2em}SLE-Net &$1.375227\times10^{7}$ &$7.077782\times10^{7}$ &$1.987178\times10^{8}$ &$4.252214\times10^{8}$ &$7.779377\times10^{8}$\hspace{2em}\\
		\hline\specialrule{0.1em}{0pt}{0.5pt}
		
	\end{tabular*}
\end{table*}

\section{Simulation Results}

In this section, the numerical results of the proposed schemes are presented and compared with traditional CI precoding. We assume that there are $100$ channel transmission blocks. In each channel transmission block, the number of symbol slots for transmission is $N_s=8000$. In each plot, we assume the total transmit power is $p_0=1$, the number of the antenna at the transmitter $M_{\mathrm{t}}$ is equal to the number of the users $K$. Both PSK modulation and QAM modulation are considered in the numerical results. Our simulation environment utilizes PyTorch version 1.13.1, and it is configured to leverage the computational power of a GeForce GTX 3090 GPU. The computational tasks on the CPU are handled by an Intel(R) Core(TM) i9-10900K CPU @ 3.70 GHz, supported by 64 GB of RAM. Additionally, MATLAB is employed for CPU-intensive operations. The following abbreviations are used throughout this section:

\begin{enumerate}
	\item ZF: traditional ZF scheme with symbol-level power normalization, where the precoding matrix of ZF is given by
	\begin{equation}
		{\bf W}_{\mathrm{ZF}} = \frac{\sqrt{p_0}}{f_{\mathrm{ZF}}}{\bf H}^{H}({\bf H}{\bf H}^{H})^{-1},
	\end{equation}
	where $f_{\mathrm{ZF}} = \left \| {\bf H}^{H}({\bf H}{\bf H}^{H})^{-1}{\bf s} \right \|_{2}$.
	\item CI-OPT: traditional optimization-based CI precoding based on ${\cal P}_{1}$ and ${\cal P}_{4}$ for PSK modulation and QAM modulation. By simplifying the problem to a quadratic optimization problem in \cite{li2018interference} and \cite{li2020interference}, it can significantly reduce run-time without performance loss.
	\item CI-CF-PSK: iterative CI precoding scheme in \cite{li2018interference}. 
	\item CI-CF-QAM: iterative CI precoding scheme in \cite{li2020interference}.
	\item SLE-CF-PSK: the iterative algorithm based on SLE for PSK modulation, introduced in Section III-A.
	\item SLE-Sub-PSK: the sub-optimal closed-form algorithm based on SLE for PSK modulation, introduced in Section III-A.
	\item SLE-CF-QAM: the iterative algorithm based on SLE for QAM modulation, introduced in Section III-B.
	\item SLE-Sub-QAM: the sub-optimal closed-form algorithm based on SLE for QAM modulation, introduced in Section III-B.
	\item SLE-Net: deep unfolding method based on SLE for the PSK modulation and QAM modulation in Section VI.
\end{enumerate}

\begin{figure}
	\centering
	\includegraphics[width=0.45\textwidth]{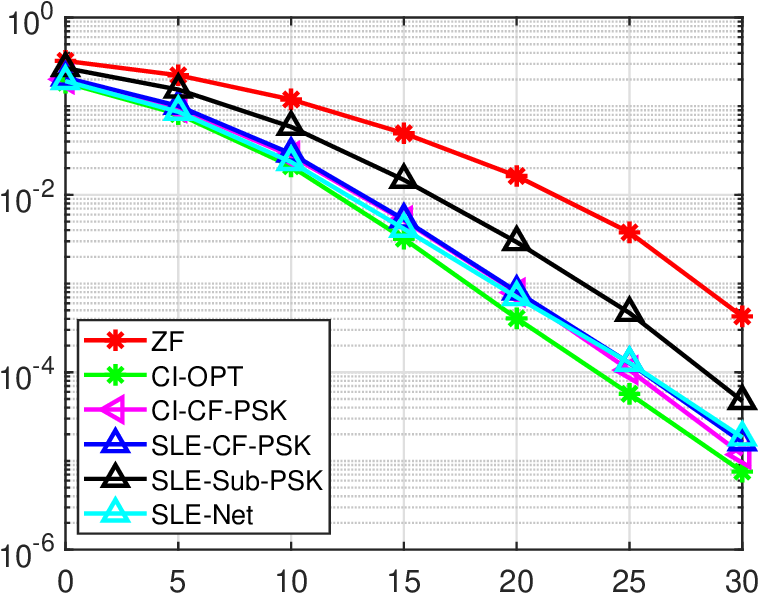}
	\captionsetup{labelformat=default,labelsep=space}
	\caption{BER v.s. SNR, QPSK, $M_{\mathrm{t}}=K=8$, $R=5$.}
	\label{qpsk_8}
\end{figure}

For QPSK modulation, we compared the bit error rate (BER) performance among different methods. As observed in Fig. \ref{qpsk_8}, the proposed SLE-Sub-PSK, SLE-CF-PSK and SLE-Net algorithms exhibit superior performance compared to the ZF when $M_{\mathrm{t}}=K=8$ and $R=5$. Both the SLE-CF-PSK and SLE-Net show a little performance loss compared to the optimal solution. the SLE-CF-PSK slightly outperforms the SLE-Net in terms of BER. Compared to the SLE-CF-PSK, the performance of SLE-Sub-PSK exhibits a significant gap. The CI-CF-PSK has a better performance than the proposed algorithms. The simulation results validate the effectiveness of the proposed SLE-based algorithm for the QPSK modulation.\par

\begin{figure}[!t]
	\centering
	\includegraphics[width=0.52\textwidth]{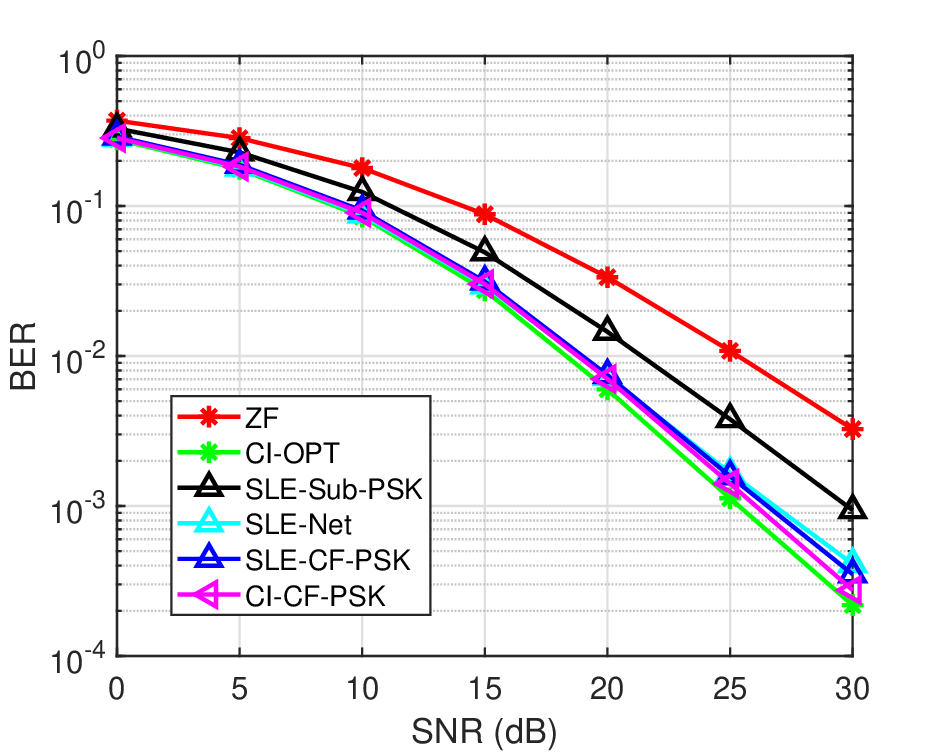}
	\captionsetup{labelformat=default,labelsep=space}
	\caption{BER v.s. SNR, 8PSK, $M_{\mathrm{t}}=K=8$, $R=5$.}
	\label{8psk_8}
\end{figure}
In Fig. \ref{8psk_8}, we show the BER performance with respect to the increasing SNR when 8PSK modulation is employed, where $M_{\mathrm{t}}=K=8$ and $R=5$. Similarly, the proposed SLE-Sub-PSK, SLE-CF-PSK and SLE-Net algorithms exhibit superior performance compared to the ZF. The simulation result validate the effectiveness of the algorithms for the 8PSK modulation scheme.

\begin{figure}
	\centering
	\includegraphics[width=0.45\textwidth]{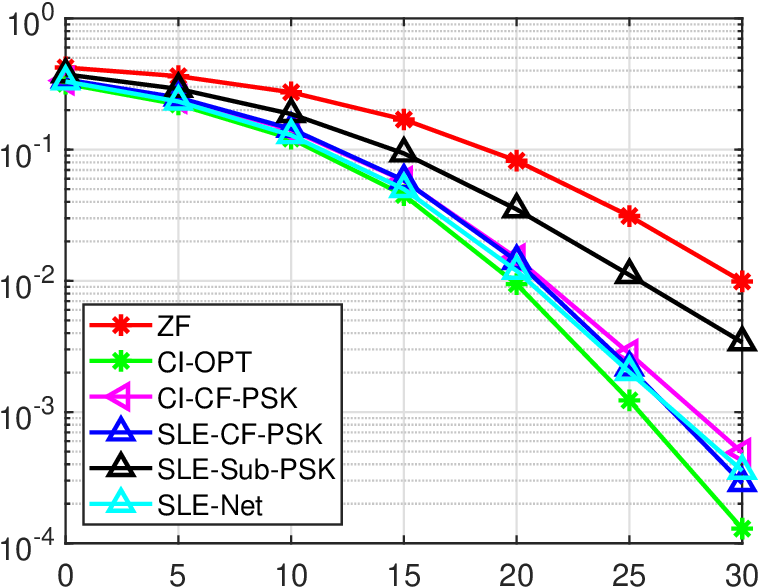}
	\captionsetup{labelformat=default,labelsep=space}
	\caption{BER v.s. SNR, 8PSK, $M_{\mathrm{t}}=K=12$, $R=5$.}
	\label{8psk_12}
\end{figure}

In Fig. \ref{8psk_12}, we consider the case of $M_{\mathrm{t}}=K=12$ and $R=5$ when 8PSK modulation is employed. The relationship of BER performance for the algorithms is not changed, and the curves exhibit the same trend compared to the case of $M_{\mathrm{t}}=K=8$.

\begin{figure}
	\centering
	\includegraphics[width=0.45\textwidth]{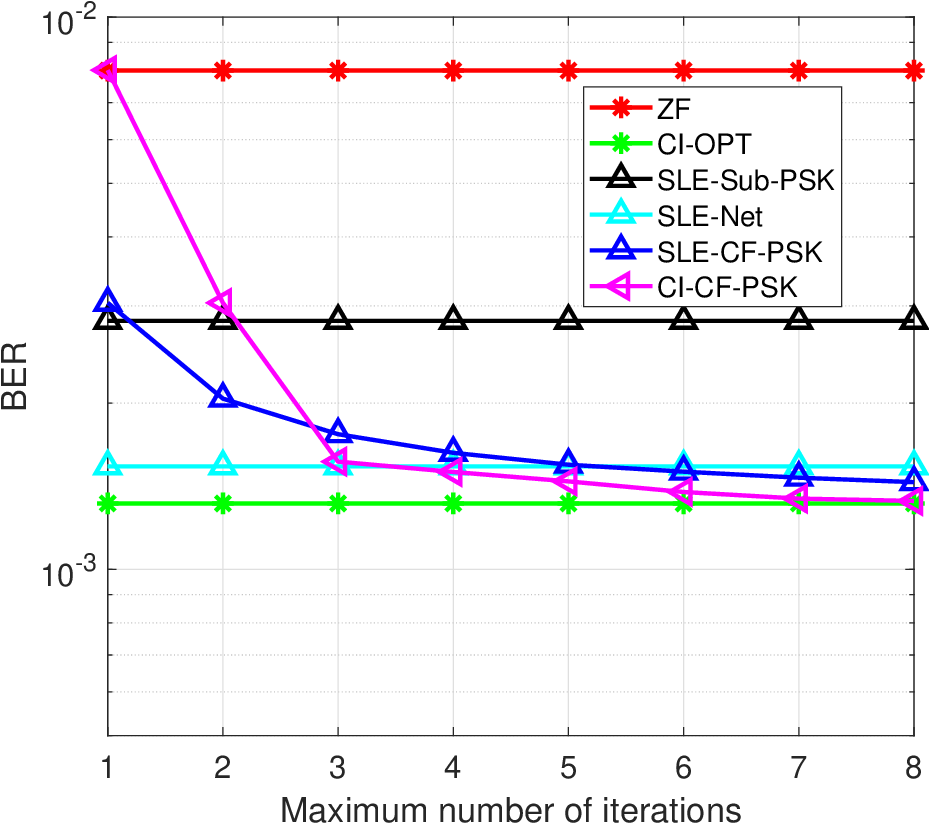}
	\captionsetup{labelformat=default,labelsep=space}
	\caption{BER v.s. Maximum number iterations, 8PSK, $M_{\mathrm{t}}=K=8$, SNR$=25$ dB.}
	\label{itera_vs_num_psk}
\end{figure}
In Fig. \ref{itera_vs_num_psk}, we illustrate the relationship between BER performance and the number of iterations. When $M_{\mathrm{t}}=K=8$ and SNR$=25$ dB, we observe that the SLE iterative algorithm gradually stabilizes its BER performance as the iteration number increasing for 8PSK, demonstrating a favorable convergence speed. At the same time, we can conclude that the SLE iterative algorithm cannot reach the optimal performance for PSK modulation.

\begin{figure}[hbt]
	\centering
	\includegraphics[width=0.45\textwidth]{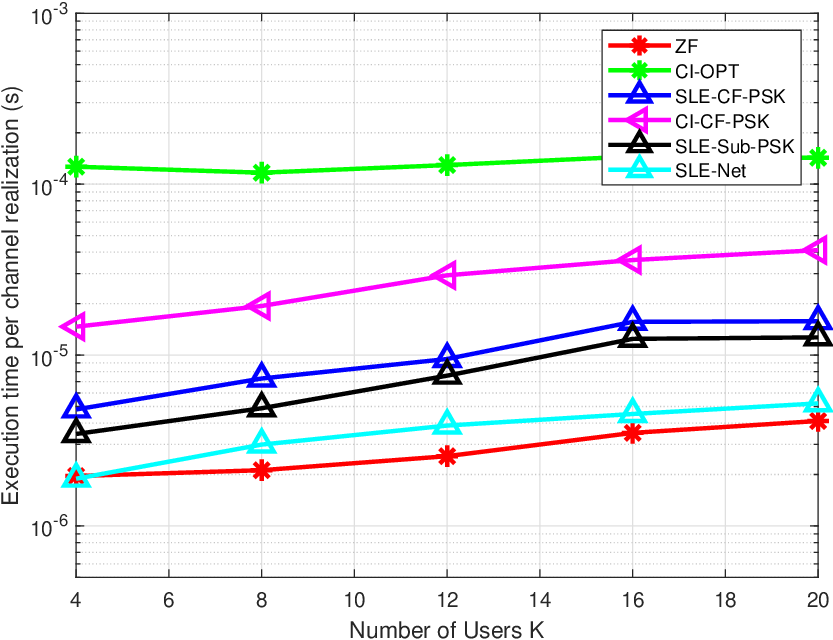}
	\captionsetup{labelformat=default,labelsep=space}
	\caption{Execution time required for different schemes, 8PSK, $M_{\mathrm{t}}=K$, $R=5$.}
	\label{time_8psk}
\end{figure}
In Fig. \ref{time_8psk}, we analyze the average execution time of all algorithms within a symbol slot when using the 8PSK modulation. It is observed that with an increasing the number of users, the execution time of all algorithms increases. Simultaneously, we conclude that the ZF has the lowest complexity. The CI-OPT has the maximized execution time. 
The execution time of SLE-CF-PSK is less than CI-CF-PSK, proving the effectiveness of the SLE. The time complexity of SLE-Net is lower than SLE-CF-PSK. This is attributed to the parallel processing capability of GPUs. 

\begin{figure}
	\centering
	\includegraphics[width=0.45\textwidth]{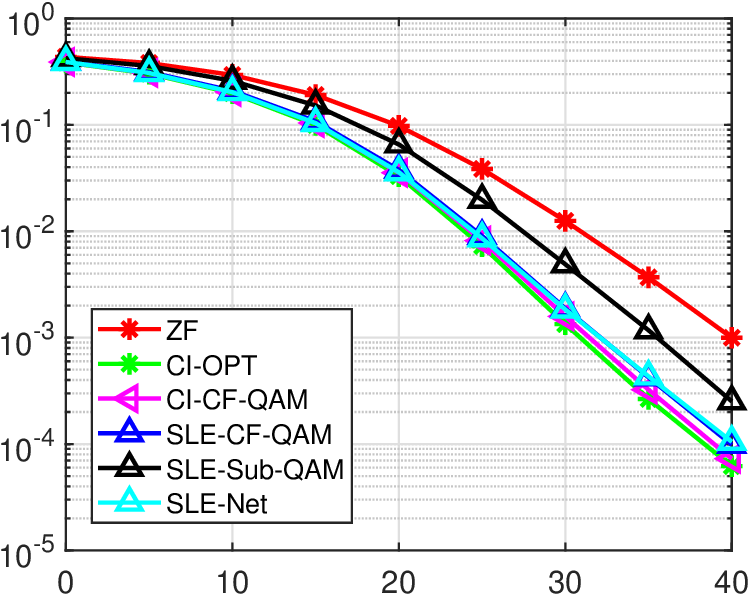}
	\captionsetup{labelformat=default,labelsep=space}
	\caption{BER v.s. SNR, 16QAM, $M_{\mathrm{t}}=K=12$, $R=5$.}
	\label{ber_qam}
\end{figure}

In Fig. \ref{ber_qam}, we show the BER performance with respect to the increasing SNR when 16QAM modulation is employed, where $M_{\mathrm{t}}=K=12$ and $R=5$. All SLE-based algorithms consistently outperform ZF. Furthermore, the performance of SLE-CF-QAM and SLE-Net has only slight degradation compared to the optimal solution.

\begin{figure}
	\centering
	\includegraphics[width=0.47\textwidth]{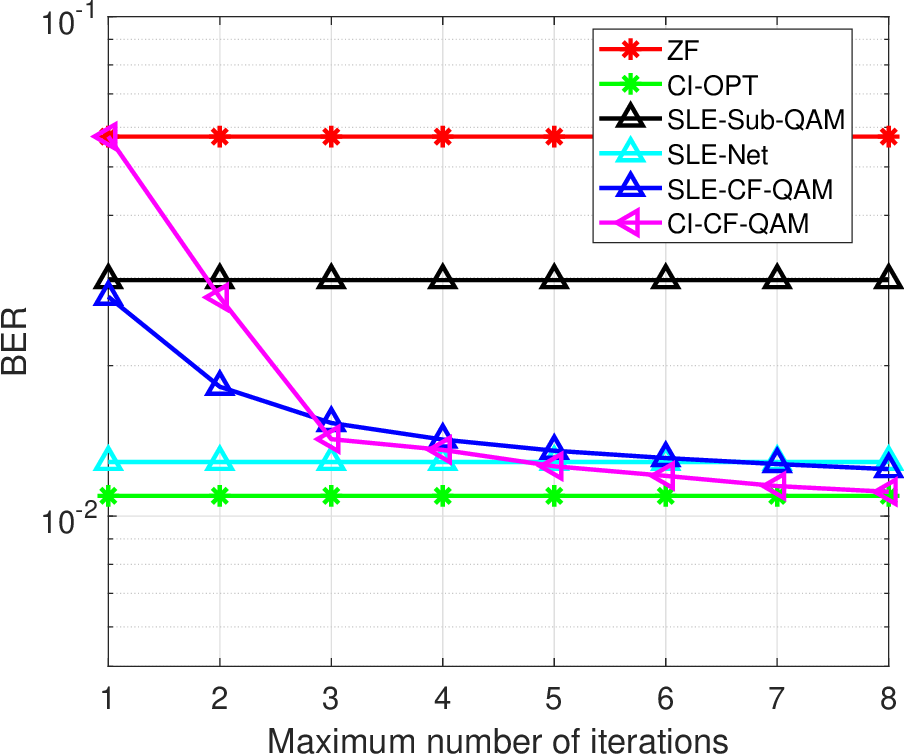}
	\captionsetup{labelformat=default,labelsep=space}
	\caption{BER v.s. Maximum number iterations, 16QAM, $M_{\mathrm{t}}=K=12$, SNR$=25$ dB.}
	\label{itera_vs_num_qam}
\end{figure}
In Fig. \ref{itera_vs_num_qam}, we show the BER performance with respect to the increasing iteration counts when 16QAM modulation is employed, where $M_{\mathrm{t}}=K=12$ and SNR $=25$ dB. The SLE-CF-QAM alse has a favorable convergence rate. The results validate the effectiveness of the proposed algorithms for the 16QAM modulation. At the same time, we can conclude that the SLE iterative algorithm cannot reach the optimal performance for QAM modulation.

\section{Conclusion}

In this paper, we propose a SLE algorithm for both the PSK modulation and QAM modulation. Within a transmission block, we exploit the relationships between the different symbol slots to extrapolate the symbol-scaling factors. By leveraging the existing solution structure of CI-SLP design, we finally obtain the precoding matrix. We further use the deep unfolding to unfold the iteration algorithm into the layer-wise structure and propose the SLE-Net. The SLE-Net is able to adapt both PSK modulation and QAM modulation without changing its network structure and loss function. The SLE-Net leverages the parallel processing capability of GPUs, which can further reduce the time complexity of algorithm. Extensive simulation results verified that the proposed algorithms have superiority in complexity reduction with an acceptable performance loss compared with the traditional SLP schemes, which ensures that SLP can be used in practical communication systems.

\bibliographystyle{IEEEtran}
\bibliography{citation}

\end{document}